\documentclass[%
reprint,
amsmath,amssymb,
aps,
prl,
]{revtex4-1}
\usepackage{graphicx}
\usepackage{dcolumn}
\usepackage{bm}
\usepackage{siunitx}
\DeclareSIUnit\gauss{G}
\usepackage{braket}
\usepackage[dvipsnames]{xcolor}
\usepackage{hyperref}
\usepackage{tabularx}

\newcommand{\mIK}{m_{I_\mathrm{K}}}
\newcommand{\mINa}{m_{I_\mathrm{Na}}}
\newcommand{\mJ}{m_J}
\begin{document}

\title{Resonant dipolar collisions of ultracold molecules induced by microwave dressing}

\author{Zoe Z. Yan$^1$, Jee Woo Park$^2$, Yiqi Ni$^1$, Huanqian Loh$^3$, Sebastian Will$^4$, Tijs Karman$^5$, and Martin Zwierlein$^1$}

\affiliation{
 $^1$MIT-Harvard Center for Ultracold Atoms, Research Laboratory of Electronics, and Department of Physics,
 Massachusetts Institute of Technology, Cambridge, Massachusetts 02139, USA\\
 $^2$Department of Physics, Pohang University of Science and Technology, Pohang 37673, Korea\\
 $^3$Department of Physics and Centre for Quantum Technologies, National University of Singapore, 117543, Singapore\\
 $^4$Department of Physics, Columbia University, New York 10027, USA\\
 $^5$ITAMP, Harvard-Smithsonian Center for Astrophysics, Cambridge, Massachusetts 02138, USA
}%

\date{\today}
\begin{abstract}
We demonstrate microwave dressing on ultracold, fermionic ${}^{23}$Na${}^{40}$K ground-state molecules and observe resonant dipolar collisions with cross sections exceeding three times the $s$-wave unitarity limit.
The origin of these collisions is the resonant alignment of the approaching molecules' dipoles along the intermolecular axis, which leads to strong attraction. We explain our observations with a conceptually simple two-state picture based on the Condon approximation. 
Furthermore, we perform coupled-channels calculations that agree well with the experimentally observed collision rates.
While collisions are observed here as laser-induced loss, microwave dressing on chemically stable molecules trapped in box potentials may enable the creation of strongly interacting dipolar gases of molecules. 
\end{abstract}

\pacs{Valid PACS appear here}
\maketitle

%
Strong, long-range dipolar interactions turn ultracold molecules into a promising platform for simulating quantum many-body physics~\cite{Micheli2006, Buchler2007,Pupillo2008, Krems2009,Yan2013}, precision measurements of fundamental constants~\cite{Carr2009, Krems2009,Andreev2018}, quantum computation~\cite{Demille2002,Yelin2006,Park2017}, and quantum state-resolved chemistry~\cite{Krems2008,Quemener2012,Balakrishnan2016,Yang2019}. Recent years have seen the production of several species of such dipolar molecular gases in the ultracold regime~\cite{Ni2008,Danzl2010,Takekoshi2014,Molony2014,Park2015,Guo2016,Rvachov2017,Seesselberg2018,Yang2019}.
A common way to induce dipolar interactions in these systems is the application of static electric fields that align molecules in the laboratory frame. To acquire dipoles on the order of the molecule's body-frame moment $d_0$ requires fields on the order of $E \sim B_\text{rot}/d_0 \sim {\rm kV}/{\rm cm}$, where $B_\text{rot}$ is the rotational constant. The presence and strength of the static electric field can be technically inconvenient.

In contrast, weak microwave electric fields that drive rotational transitions near resonance can lead to dipole moments on the order of the maximum value $d_0$. For example, dressing between the ground and first excited rotational states of a diatomic molecule yields dipole moments as large as the transition dipole moment for the electric dipole transition, $d_0/\sqrt{3}$, and thus dipolar interactions as large as 1/3 of the maximum value at a given distance.

Induced interactions via microwave dressing of molecules is a crucial component of several proposals to create exotic states of matter in bulk~\cite{Buchler2007, Micheli2007,Bruun2008, Cooper2009, Levinsen2011} and lattice quantum gases~\cite{Wall2010, Gorshkov2011}.
In addition, standing-wave microwave fields in resonant cavities have been proposed to trap polar molecules~\cite{DeMille2004,Dunseith2015,Wright2019}.
Furthermore, engineering repulsive interactions via microwave dressing can potentially shield molecules from binary collisions~\cite{Gonzalez-Martinez2017, Buchler2007, Micheli2007, Gorshkov2008, Karman2018a, lassabliere:18}, which limit the lifetime of bulk molecular gases both in chemically reactive~\cite{Ospelkaus2010,Ye2018} and non-reactive species~\cite{Takekoshi2014,Will2016,Park2017, Guo2018,Gregory2019} in the presence of trapping light~\cite{Christianen2019}.
Despite the promise of microwave dressing, its effect on collisional properties has not been studied thus far in ultracold dipolar molecules.

\begin{figure}[t]
	\begin{center}
		\includegraphics[width=\columnwidth]{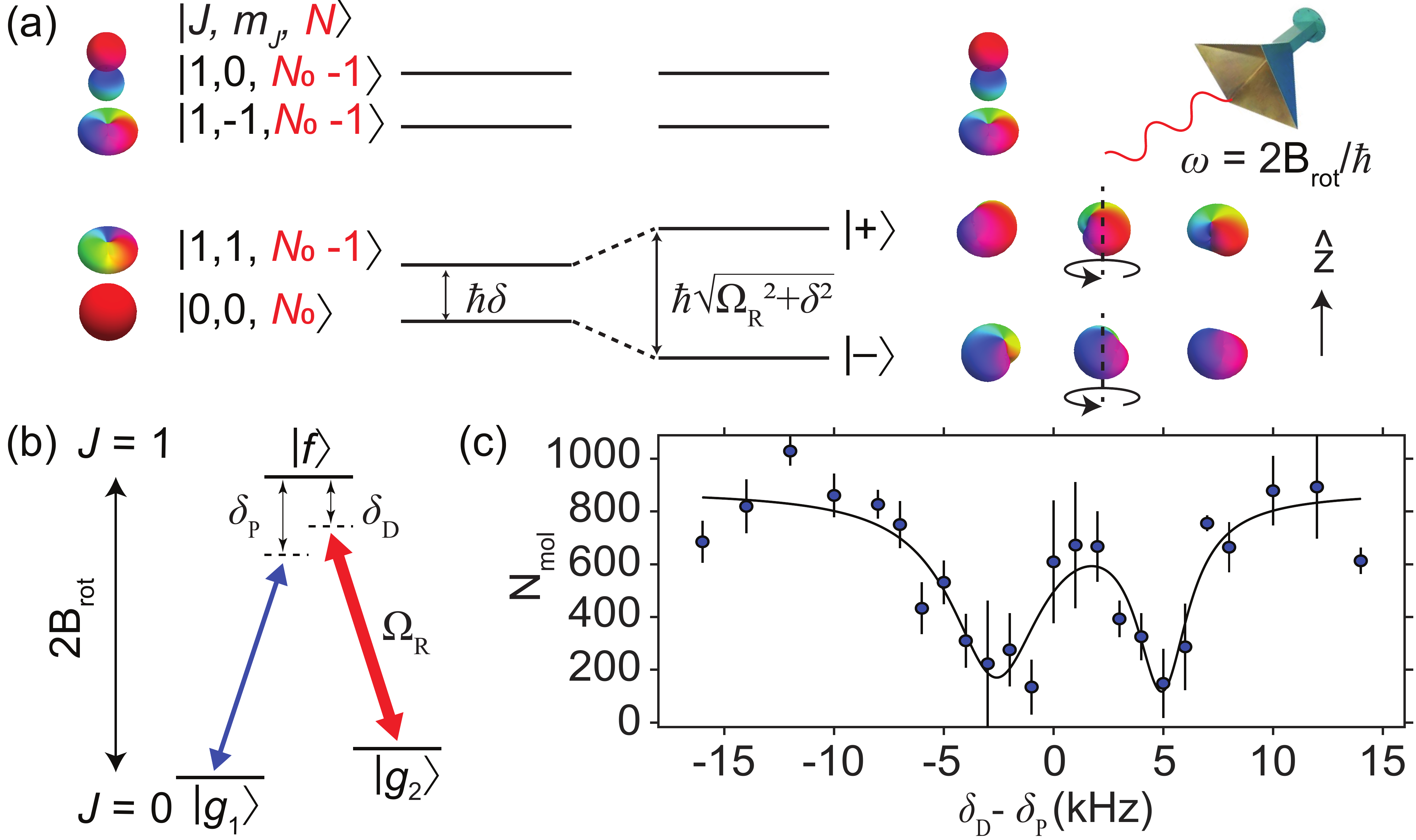}
		\caption{\label{fig:singlemolecule} Microwave dressing in ${}^{23}$Na${}^{40}$K. \textbf{(a)} Schematic energy level diagram, labeled by rotational quantum numbers $J, \mJ$ and microwave photon number $N$. Hyperfine structure is omitted for simplicity. The rotational ground state $\ket{J=0,\mJ\,{=}\,0}$ is coupled by a $\sigma^+$-polarized microwave field to the lowest energy state in the $J{=}1$ manifold, $\ket{1,1}$, resulting in dressed states $\ket{-},\ket{+}$. Higher-lying ``spectator" states (\textit{i.e.,} $\ket{1,0}$ and $\ket{1,-1}$) are not coupled by the microwaves. Molecular wavefunctions are depicted with color encoding the wavefunction's phase. 
		\textbf{(b)} Level scheme with the relevant molecular states for Autler-Townes spectroscopy. A microwave field with Rabi frequency $\Omega_\mathrm{R}/2\pi\,{=}\,$\SI{7}{\kilo\hertz} is used to address the $\ket{g_2}{\rightarrow}\ket{f}$ transition. The weaker probe microwave field has a frequency detuning $\delta_P$ that is scanned around the $\ket{g_1}{\rightarrow}\ket{f}$ transition. 
		\textbf{(c) }An Autler-Townes doublet is observed when scanning the probe microwave. The solid line shows a double Lorentzian fitted to the lineshape.
	}
	\end{center}
\end{figure}

%


In this letter, we observe strong microwave-induced interactions between fermionic $^{23}$Na$^{40}$K molecules. The employed microwaves address the transition between the ground and the first excited rotational state.
Microwave dressing enhances the probability for two molecules to reach short-range, where they can undergo light-assisted chemical reactions in the presence of the trapping laser~\cite{Christianen2019}; while ultimately this photoinduced loss can be eliminated by using repulsive box potentials~\cite{gaunt:13,mukherjee:17}, here the loss is employed as an efficient detector for the two-body collision cross section.
We find that dressing leads to resonant dipolar collisions whereby the dipoles of approaching molecules align with the intermolecular axis. This results in strong attractive interactions even for microwave detunings larger than the Rabi coupling, which we explain using a two-state model based on the Condon approximation~\cite{Julienne1996,boisseau2000reflection}. At all detunings, the collision cross sections are modelled quantitatively by coupled-channel calculations.

To start our experiment, we prepare a molecular gas in the absolute electronic, vibrational, rotational and hyperfine ground state, as described in Ref.~\cite{Park2015,Park2015a,Will2016}. In short, ultracold atomic mixtures of $^{23}$Na and $^{40}$K are confined in an optical trap at \SI{1064}{\nano\meter} and cooled to a temperature of $T \,{=}\,$\SI{560(80)}{\nano\kelvin}. 
$^{23}$Na$^{40}$K molecules are coherently associated from this sample and initialized in the lowest vibrational, rotational and hyperfine state of the ground electronic $X^1\Sigma^+$ manifold, with a peak density of $3.2(3)\,{\times}\,10^{10}$~\SI{}{\centi\meter}$^{-3}$.
Without any external electromagnetic fields, ground state $^{23}$Na$^{40}$K molecules have zero laboratory-frame dipole moment and experience no first-order dipole-dipole interaction. The dominant interaction is the background rotational van der Waals (vdW) interaction resulting from second-order dipolar coupling to the first rotationally excited state~\cite{Park2015}.
A microwave field near the resonance of the transition between the ground (rotational angular momentum quantum number $J\,{=}\,0$) and the first excited ($J\,{=}\,1$) rotational state is applied, thereby inducing a time-varying dipole moment in each molecule. The levels are spaced by the rotational splitting ${2B_\text{rot}} \,{=}\,$\SI{5.643}{\giga\hertz}, as shown in Fig.~\ref{fig:singlemolecule}(a).
Microwave dressing mixes opposite parity rotational states and imparts a significant fraction of the full dipole moment $d_0\,{=}\,2.7$\,D to the molecules. Tuned to the transition between $J\,{=}\,0$ and $J\,{=}\,1$, a resonant circularly polarized microwave field induces a dipole moment of $d_0/\sqrt{6}\,{\approx}\,1.1$\,D~\cite{Supplement,Cooper2009}, rotating with the microwave electric field.
The different hyperfine states of the first rotationally excited manifold are identified through microwave spectroscopy~\cite{Will2016}.
States are described in the nuclear-spin uncoupled basis $\ket{J,\mJ, \mINa, \mIK}$, which is an eigenbasis for the $J\,{=}\,0, \mJ\,{=}\,0$ manifold. States in $J\,{=}\,1$ are hyperfine-mixed superpositions of these basis states~\cite{Will2016}.
Therefore, a microwave field with well-defined polarization can couple the absolute ground hyperfine state, $\ket{g_1}\,{\equiv}\,\ket{0,0,-4,3/2}$, to multiple $J\,{=}\,1$ states.
Furthermore, the microwave antenna produces radiation at all polarizations: $\pi$, $\sigma^+$, and $\sigma^-$~\cite{Supplement}.

To demonstrate the presence of microwave dressing, we induce an Autler-Townes splitting of the $J\,{=}\,0\,{\rightarrow}\, J\,{=}\,1$ transition. A microwave field with Rabi frequency $\Omega_\mathrm{R}/2\pi \,{=}\, 7\,\rm kHz$ is applied on resonance with the transition between $\ket{g_2}{\equiv}\ket{0, 0, -3, 3/2}$ and $\ket{f}$, predominantly equal to $\ket{1,0,-4,3/2}$, as shown in Fig.~\ref{fig:singlemolecule}(b).
This dressing field induces a splitting of the excited state, which is probed by scanning the frequency of a weaker microwave tuned near the $\ket{g_1}{\rightarrow}\ket{f}$ transition.
We observe an Autler-Townes doublet as shown in Fig.~\ref{fig:singlemolecule}(c), demonstrating that the $\ket{g_1}$ state is only depleted by the probe field when it is tuned to the dressed resonances.

\begin{figure}[t]
	\begin{center}
		\includegraphics[width=\columnwidth]{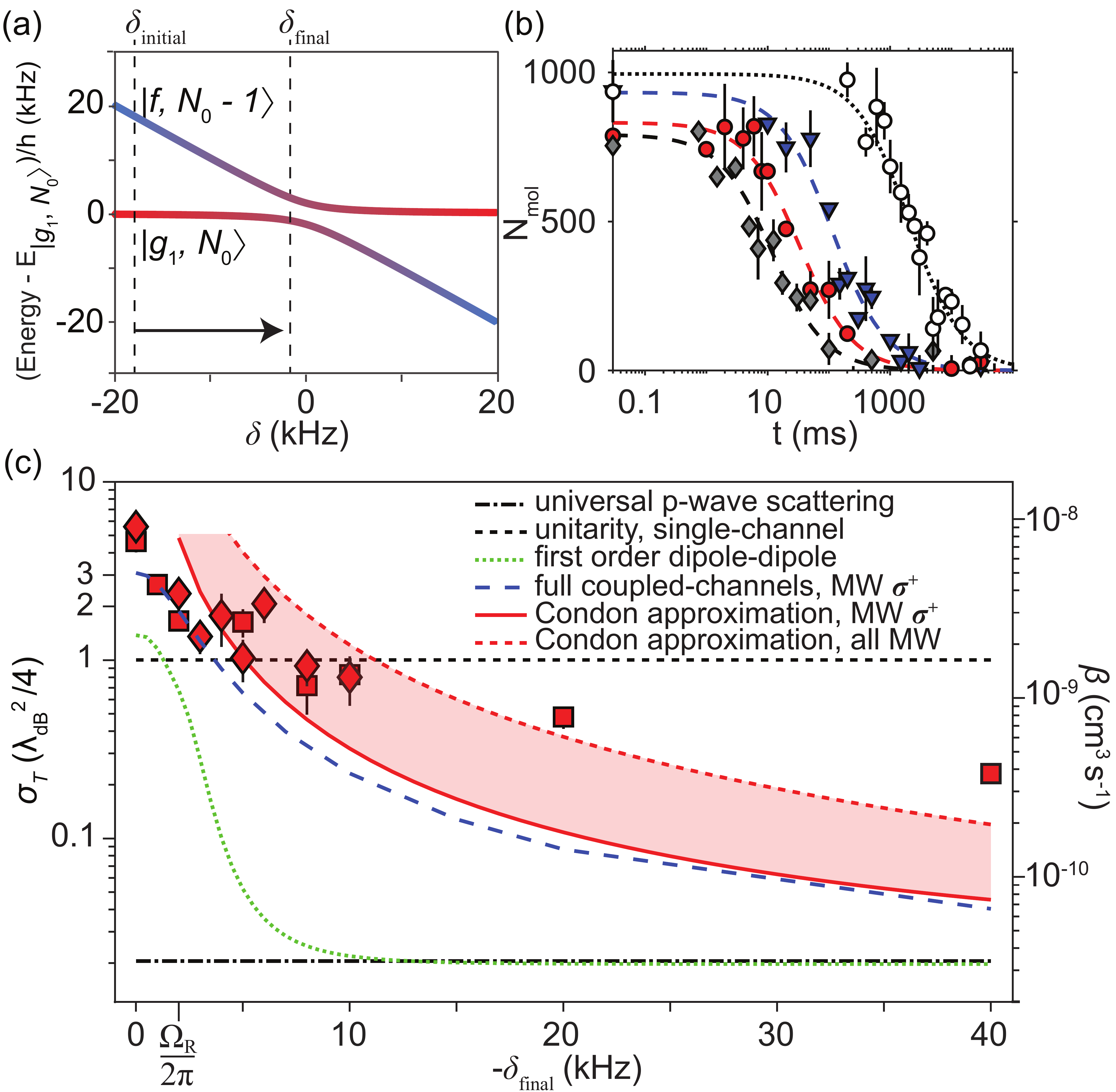}
		\caption{\label{fig:detuning} 
			Observation of resonant dipolar collisions between dressed molecules. 
			\textbf{(a)}
			Dressed energies as a function of the microwave frequency detuning $\delta$ from the $\ket{g_1}{\rightarrow}\ket{f}$ transition, at \SI{129}{\gauss}. $\delta$ is swept from far-off-resonance (typically \SI{12}{\kilo\hertz} below $\delta_\mathrm{final}$) to a final detuning at a rate of \SI{3}{\kilo\hertz\per\milli\second} and held for a varying hold time $t$.
			\textbf{(b)}
			Evolution of the molecule number under microwave dressing with $\Omega_\mathrm{R}/2\pi\,{=}\,$\SI{1.7}{\kilo\hertz} for $\delta_\mathrm{final}{=}-20,-5$, and 0 kHz (in blue triangles, red circles, and grey diamonds). A lifetime curve of $\ket{g_1}$ taken without microwaves is shown as open circles.
			Dashed lines are two-body decay fits. 
			\textbf{(c)} 	Collision rates (left axis) obtained from loss coefficients (right axis) of dressed molecules as a function of $\delta_\mathrm{final}$ for $\Omega_\mathrm{R}/2\pi\,{=}\,$\SI{1.7}{\kilo\hertz} (squares) and \SI{2.4}{\kilo\hertz} (diamonds).
			The black dot-dashed line shows the universal $p$-wave loss at \SI{560}{\nano\kelvin}~\cite{Idziaszek2010}, the green dotted line includes the additional loss from first-order dipole-dipole interactions. The unitarity limit for the loss rate from a single partial wave is shown as the black dashed line.
			The blue dashed line shows the coupled-channels prediction, assuming pure $\sigma^+$ microwave polarization, and the red line depicts the rate given by the Condon approximation. Including all microwave polarizations, the Condon approximation increases to the red dashed line.
		}
	\end{center}
\end{figure}

We find that microwave dressing dramatically enhances molecular interactions. 
Although ${}^{23}$Na${}^{40}$K should not experience two-body collisional loss in its electronic and vibrational ground state, the trapping laser at \SI{1064}{\nano\meter} leads to photoinduced loss at short range~\cite{Christianen2019}. We employ this loss mechanism as a probe for microwave-induced two-body collisions.
To start, the dressing microwave field is first applied with a frequency far below the lowest rotational resonance, the $\ket{g_1}{\rightarrow}\ket{f}$ transition. 
Here, and for the remainder of the paper, $\ket{f}$ represents the lowest energy $J\,{=}\,1$ state, which has predominantly $\ket{1,1,-4,3/2}$ character.
The frequency is then swept adiabatically from the initial detuning $\delta_\mathrm{initial}$, where the molecule in the lower dressed eigenstate $\ket{-}$ has predominantly $\ket{g_1}$ character, to a detuning $\delta_\mathrm{final}$ near or on the dressed resonance [see Fig.~\ref{fig:detuning}(a)].
Fig.~\ref{fig:singlemolecule}(a) depicts the dressed eigenstate $\ket{-}$, which is a superposition of the states $\ket{g_1}$ and $\Ket{f}$.
The red-detuned microwaves avoid driving other hyperfine transitions during the sweep (the ``spectator" states of Fig.~\ref{fig:singlemolecule}(a)); the next higher $J\,{=}\,1$ state lies 27 kHz above the $\ket{f}$ state.
The microwave field is held at its final detuning for a varying amount of time, allowing collisions to occur, before the detuning is swept back to $\delta_\mathrm{initial}$ and the remaining $\ket{g_1}$ molecules are imaged.

We observe the evolution of the molecule number in $\ket{g_1}$ as a function of hold time to extract the loss rate of the ensemble; examples for certain detunings are shown in Fig.~\ref{fig:detuning}(b).
The loss curves are fit to a two-body decay model, where the density $n(t)$ as a function of time obeys $n(t)=n_0/(1+\beta n_0 t)$.
Here, $n_0$ is the initial average molecule density and $\beta$ is the two-body loss coefficient.
The microwave dressing shortens the sample lifetime by orders of magnitude, compared to the $\sim$\SI{3}{\second} lifetime in the absence of microwaves~\cite{Park2015,Will2016}.
Since both molecules involved in the collision will be lost, the loss rate is related to the two-body scattering cross section $\sigma$ by $\beta = 2 \left<\sigma v\right>$, where $\left<\dots\right>$ denotes the ensemble average and $v$ the relative velocity of colliding molecules. We therefore define a thermally averaged scattering cross section $\sigma_{\rm T} \,{=}\, \beta/2\left<v\right>$, with $\langle v \rangle \,{=}\, \sqrt{8k_\mathrm{B}T/\pi \mu}$ the average relative velocity and $\mu \,{=}\, (m_{\rm Na}+m_{\rm K})/2$ the reduced mass.

Figure~\ref{fig:detuning}(c) shows the measured collision cross section (red data points, left axis) and associated loss rate (right axis) as a function of microwave detuning.
The resonant scattering rate is an order of magnitude larger than rates found in previous experiments~\cite{Ni2010,Guo2018, Gregory2019}.
Away from resonance, the scattering cross section is reduced but remains orders of magnitude larger than that of both the bare $\ket{g_1}$ and $\ket{f}$ states in the absence of microwaves. The bare states feature loss rates of only $\beta^\mathrm{(bare)}\,{=}\,2{\times}10^{-11}$~cm$^3$s$^{-1}$~\cite{Park2015,Will2016}, close to the universal loss rate of $\beta^\mathrm{(universal)}\,{=}\,3{\times}10^{-11}$~cm$^3$s$^{-1}$~\cite{Idziaszek2010}, which reflects the loss when the molecules only experience vdW interactions under $p$-wave collisions. 

To emphasize how strongly microwave dressing can modify interactions, the comparison to the $s$-wave unitarity limit is useful. A single partial wave contributes at most the unitarity limit, $\sigma^{(\rm unitarity)} = \lambda_\mathrm{dB}^2/4$, limited by the de Broglie wavelength $\lambda_\mathrm{dB} = \sqrt{2\pi\hbar^2/\mu k_\mathrm{B} T}$, corresponding to a loss rate $\beta^{(\mathrm{unitarity})} \,{=}\,2\hbar\lambda_\mathrm{dB}/\mu \,{=}\, 1.7\times10^{-9}$~cm$^3$s$^{-1}$.
For ultracold bosons that undergo only $s$-wave collisions, 
$\sigma^{(\rm unitarity)}$ and $\beta^{(\mathrm{unitarity})}$ are upper limits to the collisional cross section and loss rate, respectively. For ultracold fermions such as ${}^{23}$Na${}^{40}$K, 
one might expect the $p$-wave centrifugal barrier to prevent molecules from reaching short-range and thus reduce losses, but the dipole-dipole interaction suppresses this barrier.
For dressing on a $\sigma^+$ resonance,
the first-order dipole-dipole interaction is attractive for $M_L\,{=}\,\pm 1$, where $M_L$ is the projection of the molecules' relative angular momentum~\cite{Supplement},
leading to a $p$-wave loss rate that is at most twice the unitarity limit.

In the remainder of the paper we explain the origins of the dressing-induced collisions. We first consider a simple description where the molecules in the $\ket{-}$ state only experience background vdW interactions and the first-order dipole-dipole interaction, neglecting all ``spectator states" and the upper dressed state $\ket{+}$.
In this approximation, the molecular dipole moments always align with the rotating electric field.
The resulting loss curve, shown as the green dotted line in Fig.~\ref{fig:detuning}(c), is comparable to the unitarity limit near resonance.
Away from resonance where $|\delta_{\rm{final}}| \,{\gg}\,\Omega_{\rm{R}}/2\pi$, microwave dressing induces a negligible dipole moment and the first order approximation to the collision rate rapidly decreases to the universal limit.
This disagrees with the experimental loss rates, which remain an order of magnitude higher than the bare rate without microwave dressing even for detunings greater than $\Omega_R$.
Thus, the first order dipole-dipole effect is insufficient to explain the enhanced collision rates.

\begin{figure}[t]
	\begin{center}
		\includegraphics[width=\columnwidth]{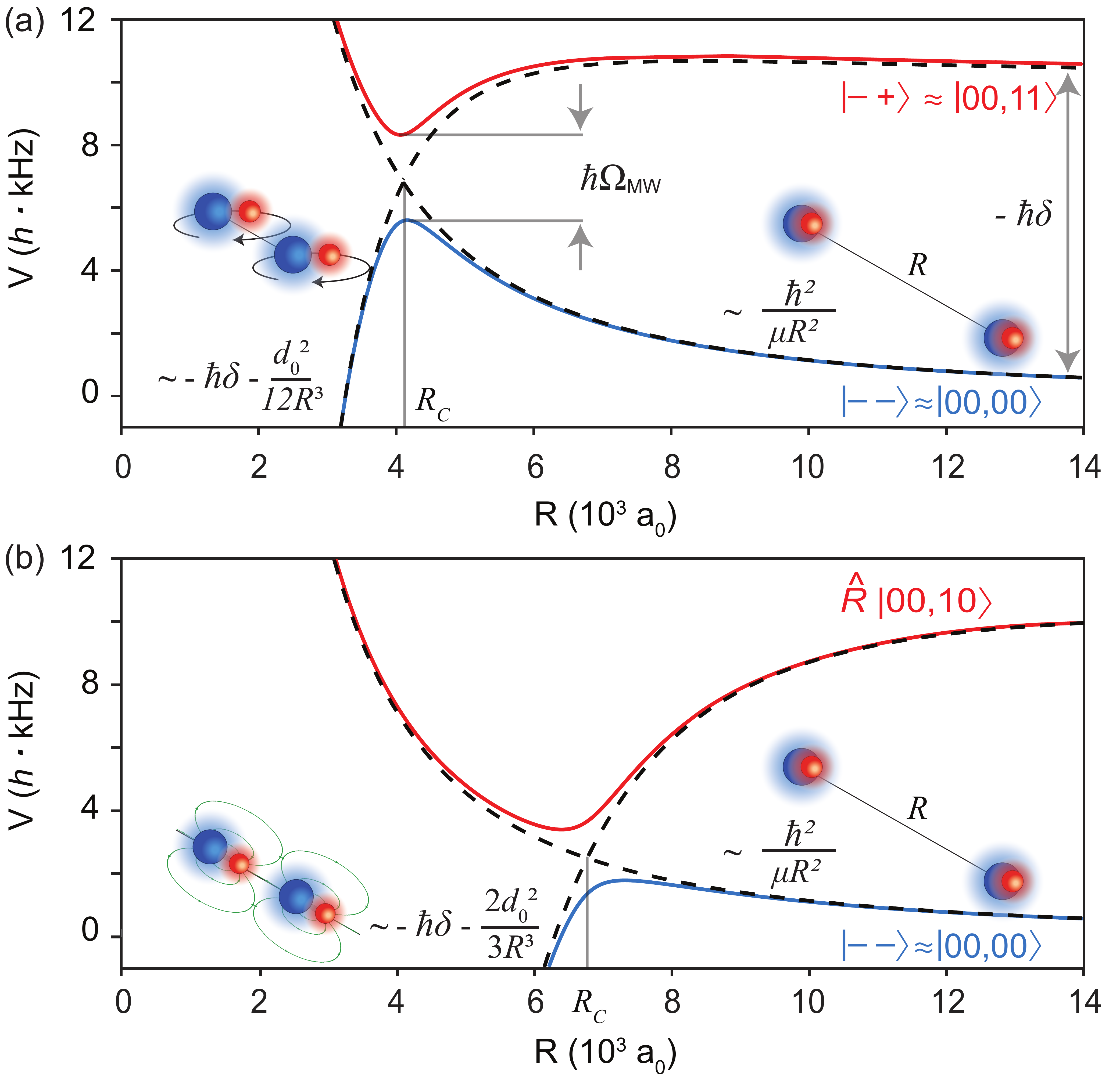}
		\caption{\label{fig:barriers} 
			Two-state picture of dipolar interactions between microwave-dressed molecules, 
			valid for detunings larger than the Rabi coupling, and shown here for $M_L\,{=}\,1$. \textbf{(a)} Interaction potentials for molecules in the $\ket{--}$ and $\ket{-+}$ states. An effective microwave Rabi coupling between the branches causes an avoided crossing at $R_\mathrm{C}$.
			Excluding spectator states, the molecules remain aligned with the microwave field. 
			$\Omega_{\rm MW}$ is the effective Rabi frequency between the two states, proportional to $\Omega_\mathrm{R}$~\cite{Supplement}.
			\textbf{(b)} The same as (a) but including spectator states: the relevant excited potential comes from two molecules in the $\hat{\mathcal{R}} \left( |00\rangle|10\rangle + |10\rangle|00\rangle\right)/\sqrt{2}$ state~\cite{Supplement}, which represents the molecules aligning at short range along the intermolecular axis $\hat{R}$. Here, molecules experience strong, resonant dipole-dipole interactions.
		}
	\end{center}
\end{figure}

Next, we also consider contributions from the upper dressed state $\ket{+}$, restricting the two-molecule basis to $\ket{--}$ and $(\ket{+-}+\ket{-+}){/}\sqrt{2}$, written as $\ket{-+}$ for convenience in the remainder of the paper. This approximation is valid at detunings greater than $\Omega_R$, when accounting for only $\sigma^+$ microwave polarization and neglecting the presence of ``spectator states'' $\left|J\,{=}\,1,m_J\,{=}\,0\right>$ and $\left|1,-1\right>$. 
Neglecting these spectators, the dipoles can only ever point in the direction of the rotating microwave electric field, \textit{i.e.,} in the $x$-$y$ plane, as they approach each other at close range.
The interactions will thus be repulsive if molecules meet along the $z$-direction ($M_L\,{=}\,0$) and attractive if they meet in the $x$-$y$ plane, i.e. for $M_L\,{=\pm}\, 1$.
One might therefore expect a maximum $p$-wave cross section of at most \textit{twice} the unitarity limit corresponding to the two attractive $M_L\,{=\pm}\, 1$ channels for thermal energies far greater than the barrier height.
For red detunings exceeding the Rabi frequency, the adiabatic potential curves for $L=1$, $M_L=1$ display an avoided crossing between the incoming centrifugal potential ${\sim} \hbar^2/\mu R^2$, with negligible dipolar interaction, and the attractive potential ${\sim}{-}h\delta_\mathrm{final}{-}d_0^2 /12R^3{+}\hbar^2/\mu R^2$ [see Fig.~\ref{fig:barriers}(a)], 
 corresponding to the time-averaged dipolar attraction of two classical rotating dipoles of strength $d_0/\sqrt{6}$ approaching in the plane of rotation.
The diabatic potentials cross at the Condon point $R_\mathrm{C}$, and an effective Rabi coupling leads to a $p$-wave barrier $V_b\,{\sim}\,\hbar^2/\mu R_\mathrm{C}^2 \,{\propto}\, |\delta_\mathrm{final}|^{2/3}$.
Incoming molecules can reach this barrier, entering short range, leading to high scattering rates even for off-resonant dressing.

However, a quantitative comparison to the data requires treating the ``spectators'' $\left|J\,{=}\,1,m_J\,{=}\,0\right>$ and $\left|1,-1\right>$, as they are sufficiently close in energy. These ``spectators" enable the molecules to reorient so that their dipoles point head to tail [see Fig.~\ref{fig:barriers}(b)], leading to resonant dipole-dipole interactions. This occurs when the dipolar energy overcomes the energy difference between the dressed incoming state and the state of attractively interacting molecules, or classically, when the electric field applied by one molecule on the other exceeds the electric field of the applied microwaves. Thus at short range the interaction between two microwave-dressed molecules incoming in the lowest internal state $\ket{--}$ will always be attractive regardless along which direction the molecules meet, \textit{i.e.,} for all three $M_L$ components, giving a potential $ {\sim}\,{-}\,2 d_0^2/3R^3$. This resonant dipolar collision leads to $p$-wave loss as high as \textit{three times} the unitarity limit.
Even faster losses require inclusion of higher partial waves, $L\,{>}\,1$.
Compared to the spectator-free case of Fig.~\ref{fig:barriers}(a), the barrier height is shifted down and the Condon point is moved outwards.
Losses can be analytically derived for detunings exceeding the Rabi frequency by the reflection approximation for the Franck-Condon overlap \cite{Julienne1996,burnett1996laser,boisseau2000reflection},
\begin{align}
	\beta = \frac{16\pi^2}{9\hbar} \left(\frac{\Omega_{\rm{R}}}{\delta}\right)^2\, d_0^2 \,\langle j_l(k R_\mathrm{C})^2 \rangle  
\end{align}
where angular brackets indicate averaging over the thermal velocity distribution, $k$ is the collision wavevector, and $j_l$ is the spherical Bessel function of the first kind. 
The resulting approximation is shown as the solid red line in Fig.~\ref{fig:detuning}(c).

For a full model of the observed loss curves we employ coupled-channel (c.c.)\ calculations~\cite{Supplement} [see the dashed blue line of Fig.~\ref{fig:detuning}(c)].
The molecules are represented as rigid rotors with hyperfine structure that interact through dipole-dipole interactions and undergo photoinduced loss at short range, modeled by an absorbing boundary condition.
The scattering calculations capture both the high loss rate on resonance, exceeding three times the $s$-wave unitarity limit, and the slow decrease with detuning: even at $|\delta_\mathrm{final}|{\gg}\, \Omega_\mathrm{R}/2\pi\,{\approx}\,\SI{2}{\kilo\hertz}$, the loss is significantly higher than the universal loss rate, obtained without microwave dressing.
However, the experimentally observed loss decreases even more slowly with detuning than for c.c.\ calculations that include only $\sigma^+$ polarization. We attribute this to the $\pi$ and $\sigma^-$ components of the microwave field \cite{Supplement}. On the $\sigma^+$ resonance, these field components address far-detuned hyperfine transitions,
and their effect can be neglected.
Away from resonance, however, these field components should contribute comparably and hence double the effective Rabi frequency~\cite{Supplement}.
This effect is also incorporated into the Condon approximation, resulting in the red dashed line in Fig.~\ref{fig:detuning}(c). The adjusted Condon approximation matches the experiment at higher detunings.

\begin{figure}
	\begin{center}

		\includegraphics[width=\columnwidth]{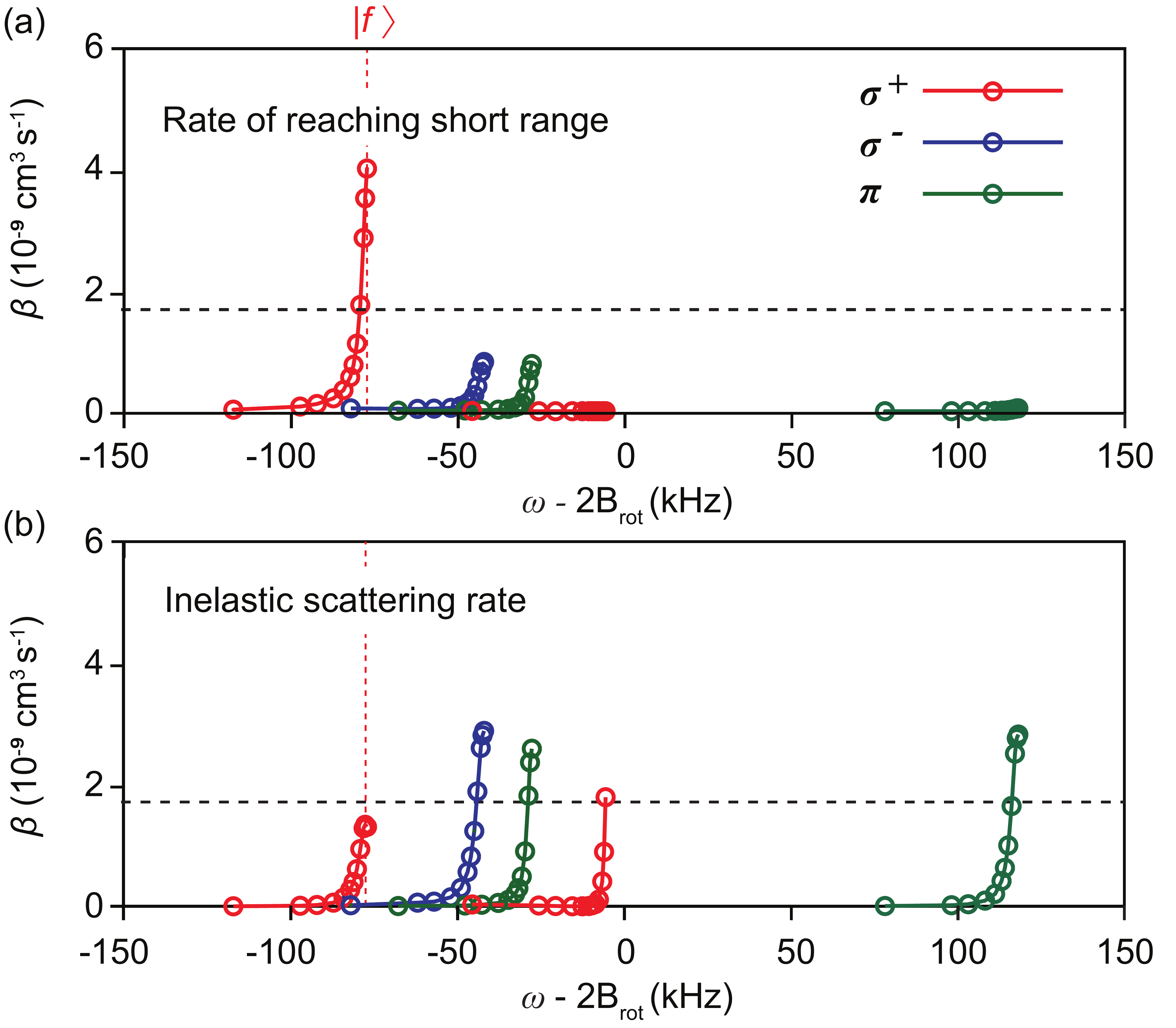}
		\caption{ \label{fig:spectrum}
		 Hyperfine-state-dependent interactions, calculated at \SI{129}{\gauss}. 
         \textbf{(a)} 
         The short-range loss, which occurs when the incoming molecules reach the absorptive boundary condition that models photoinduced loss, is strongest for the lowest hyperfine state $\ket{f}$, whereas excited hyperfine states have larger inelastic loss (shown in \textbf{(b)}) due to transitions to different field-dressed levels and hyperfine states. The single-channel unitarity limit is shown as the dotted line.
	}
	\end{center}
\end{figure}

The dressing-induced collisions are affected by hyperfine interactions that shift the ``spectator states'' relative to the state used for microwave dressing.
Recall that the maximum strength of resonant dipolar interactions is given by the full transition dipole matrix element of the dressed transition. This requires reorientation of the molecules along their intermolecular axis and thus inclusion of the relevant $\ket{J,\mJ}$ states, typically split by the hyperfine interaction. 
Hence, resonant interactions take full effect when the dipole-dipole interaction is large compared to the hyperfine splittings.

Though here the microwave dressing was on the lowest $J\,{=}\,1$ state, the choice to dress on a higher hyperfine state would affect the induced collision rates. 
Potential curves for dressing on states higher than $\ket{f}$ exhibit many crossings rather than approaching an isolated attractive resonant dipole-dipole potential \cite{Supplement}, leading to slower scattering rates for reaching short range [Fig.~\ref{fig:spectrum}(a)].
Additionally, nonadiabatic transitions into lower-lying hyperfine states may increase the inelastic losses of the reflected flux to hyperfine states other than the initial channel, compared to the case of dressing on $\ket{f}$ [Fig.~\ref{fig:spectrum}(b)].

State-dependent resonant dipolar interactions induced by microwave dressing, found here, will enrich applications of polar molecules in quantum computation and simulation of many-body physics~\cite{Micheli2006, Buchler2007,Pupillo2008,Krems2009}.
The characteristic range $R_\mathrm{C}$ where the resonant dipolar collision occurs is directly controlled by the microwave detuning. This range can easily reach the typical spacings in optical lattices, $\sim$~\SI{500}{\nano\meter}, enabling dipolar exchange energies to dominate over all other relevant energy scales in the system.
Here we observed dressing through collisional losses, but under appropriate conditions (\textit{e.g.}, molecules trapped in a repulsive optical ``box" potential) the short-range photo-induced losses should not occur. In such situations, microwave and electric field dressing can lead to strong elastic scattering, offering a powerful technique to tune intermolecular interactions.
Understanding and harnessing such interactions in ultracold polar molecules will be crucial for the creation of novel phases of matter, including topological superfluidity~\cite{Cooper2009}.

We would like to thank Alexey Gorshkov for stimulating discussions.
This work was supported by the NSF, AFOSR, ARO, an AFOSR MURI on ``Exotic Phases of Matter," the David and Lucile Packard Foundation, the Vannevar Bush Faculty Fellowship, and the Gordon and Betty Moore Foundation through grant GBMF5279. Z.Z.Y. acknowledges support from the NSF GRFP. T.K. acknowledges support from NWO Rubicon grant 019.172EN.007 and an NSF grant to ITAMP. S.W. acknowledges support by the Alfred P. Sloan Foundation. H.L. acknowledges support from the National Research Foundation Singapore.

\nocite{Will2016,Karman2018a,janssen:13,Park2015,aldegunde:17}

\bibliography{MWRefs}

\begin{thebibliography}{54}%
\makeatletter
\providecommand \@ifxundefined [1]{%
 \@ifx{#1\undefined}
}%
\providecommand \@ifnum [1]{%
 \ifnum #1\expandafter \@firstoftwo
 \else \expandafter \@secondoftwo
 \fi
}%
\providecommand \@ifx [1]{%
 \ifx #1\expandafter \@firstoftwo
 \else \expandafter \@secondoftwo
 \fi
}%
\providecommand \natexlab [1]{#1}%
\providecommand \enquote  [1]{``#1''}%
\providecommand \bibnamefont  [1]{#1}%
\providecommand \bibfnamefont [1]{#1}%
\providecommand \citenamefont [1]{#1}%
\providecommand \href@noop [0]{\@secondoftwo}%
\providecommand \href [0]{\begingroup \@sanitize@url \@href}%
\providecommand \@href[1]{\@@startlink{#1}\@@href}%
\providecommand \@@href[1]{\endgroup#1\@@endlink}%
\providecommand \@sanitize@url [0]{\catcode `\\12\catcode `\$12\catcode
  `\&12\catcode `\#12\catcode `\^12\catcode `\_12\catcode `\%12\relax}%
\providecommand \@@startlink[1]{}%
\providecommand \@@endlink[0]{}%
\providecommand \url  [0]{\begingroup\@sanitize@url \@url }%
\providecommand \@url [1]{\endgroup\@href {#1}{\urlprefix }}%
\providecommand \urlprefix  [0]{URL }%
\providecommand \Eprint [0]{\href }%
\providecommand \doibase [0]{http://dx.doi.org/}%
\providecommand \selectlanguage [0]{\@gobble}%
\providecommand \bibinfo  [0]{\@secondoftwo}%
\providecommand \bibfield  [0]{\@secondoftwo}%
\providecommand \translation [1]{[#1]}%
\providecommand \BibitemOpen [0]{}%
\providecommand \bibitemStop [0]{}%
\providecommand \bibitemNoStop [0]{.\EOS\space}%
\providecommand \EOS [0]{\spacefactor3000\relax}%
\providecommand \BibitemShut  [1]{\csname bibitem#1\endcsname}%
\let\auto@bib@innerbib\@empty
\bibitem [{\citenamefont {Micheli}\ \emph {et~al.}(2006)\citenamefont
  {Micheli}, \citenamefont {Brennen},\ and\ \citenamefont
  {Zoller}}]{Micheli2006}%
  \BibitemOpen
  \bibfield  {author} {\bibinfo {author} {\bibfnamefont {A.}~\bibnamefont
  {Micheli}}, \bibinfo {author} {\bibfnamefont {G.~K.}\ \bibnamefont
  {Brennen}}, \ and\ \bibinfo {author} {\bibfnamefont {P.}~\bibnamefont
  {Zoller}},\ }\href {\doibase 10.1038/nphys287} {\bibfield  {journal}
  {\bibinfo  {journal} {Nat. Phys.}\ }\textbf {\bibinfo {volume} {2}},\
  \bibinfo {pages} {341} (\bibinfo {year} {2006})}\BibitemShut {NoStop}%
\bibitem [{\citenamefont {B{\"{u}}chler}\ \emph {et~al.}(2007)\citenamefont
  {B{\"{u}}chler}, \citenamefont {Demler}, \citenamefont {Lukin}, \citenamefont
  {Micheli}, \citenamefont {Prokof'ev}, \citenamefont {Pupillo},\ and\
  \citenamefont {Zoller}}]{Buchler2007}%
  \BibitemOpen
  \bibfield  {author} {\bibinfo {author} {\bibfnamefont {H.~P.}\ \bibnamefont
  {B{\"{u}}chler}}, \bibinfo {author} {\bibfnamefont {E.}~\bibnamefont
  {Demler}}, \bibinfo {author} {\bibfnamefont {M.}~\bibnamefont {Lukin}},
  \bibinfo {author} {\bibfnamefont {A.}~\bibnamefont {Micheli}}, \bibinfo
  {author} {\bibfnamefont {N.}~\bibnamefont {Prokof'ev}}, \bibinfo {author}
  {\bibfnamefont {G.}~\bibnamefont {Pupillo}}, \ and\ \bibinfo {author}
  {\bibfnamefont {P.}~\bibnamefont {Zoller}},\ }\href {\doibase
  10.1103/PhysRevLett.98.060404} {\bibfield  {journal} {\bibinfo  {journal}
  {Phys. Rev. Lett.}\ }\textbf {\bibinfo {volume} {98}},\ \bibinfo {pages}
  {060404} (\bibinfo {year} {2007})}\BibitemShut {NoStop}%
\bibitem [{\citenamefont {Pupillo}\ \emph {et~al.}(2008)\citenamefont
  {Pupillo}, \citenamefont {Griessner}, \citenamefont {Micheli}, \citenamefont
  {Ortner}, \citenamefont {Wang},\ and\ \citenamefont {Zoller}}]{Pupillo2008}%
  \BibitemOpen
  \bibfield  {author} {\bibinfo {author} {\bibfnamefont {G.}~\bibnamefont
  {Pupillo}}, \bibinfo {author} {\bibfnamefont {A.}~\bibnamefont {Griessner}},
  \bibinfo {author} {\bibfnamefont {A.}~\bibnamefont {Micheli}}, \bibinfo
  {author} {\bibfnamefont {M.}~\bibnamefont {Ortner}}, \bibinfo {author}
  {\bibfnamefont {D.~W.}\ \bibnamefont {Wang}}, \ and\ \bibinfo {author}
  {\bibfnamefont {P.}~\bibnamefont {Zoller}},\ }\href {\doibase
  10.1103/PhysRevLett.100.050402} {\bibfield  {journal} {\bibinfo  {journal}
  {Phys. Rev. Lett.}\ }\textbf {\bibinfo {volume} {100}},\ \bibinfo {pages}
  {050402} (\bibinfo {year} {2008})}\BibitemShut {NoStop}%
\bibitem [{\citenamefont {Krems}\ \emph {et~al.}(2009)\citenamefont {Krems},
  \citenamefont {Stwalley},\ and\ \citenamefont {Friedrich}}]{Krems2009}%
  \BibitemOpen
  \bibfield  {author} {\bibinfo {author} {\bibfnamefont {R.~V.}\ \bibnamefont
  {Krems}}, \bibinfo {author} {\bibfnamefont {W.~C.}\ \bibnamefont {Stwalley}},
  \ and\ \bibinfo {author} {\bibfnamefont {B.}~\bibnamefont {Friedrich}},\
  }\href@noop {} {\emph {\bibinfo {title} {Cold Molecules: Theory, Experiment,
  Applications}}}\ (\bibinfo  {publisher} {CRC Press},\ \bibinfo {address}
  {Boca Raton},\ \bibinfo {year} {2009})\BibitemShut {NoStop}%
\bibitem [{\citenamefont {Yan}\ \emph {et~al.}(2013)\citenamefont {Yan},
  \citenamefont {Moses}, \citenamefont {Gadway}, \citenamefont {Covey},
  \citenamefont {Hazzard}, \citenamefont {Rey}, \citenamefont {Jin},\ and\
  \citenamefont {Ye}}]{Yan2013}%
  \BibitemOpen
  \bibfield  {author} {\bibinfo {author} {\bibfnamefont {B.}~\bibnamefont
  {Yan}}, \bibinfo {author} {\bibfnamefont {S.~A.}\ \bibnamefont {Moses}},
  \bibinfo {author} {\bibfnamefont {B.}~\bibnamefont {Gadway}}, \bibinfo
  {author} {\bibfnamefont {J.~P.}\ \bibnamefont {Covey}}, \bibinfo {author}
  {\bibfnamefont {K.~R.}\ \bibnamefont {Hazzard}}, \bibinfo {author}
  {\bibfnamefont {A.~M.}\ \bibnamefont {Rey}}, \bibinfo {author} {\bibfnamefont
  {D.~S.}\ \bibnamefont {Jin}}, \ and\ \bibinfo {author} {\bibfnamefont
  {J.}~\bibnamefont {Ye}},\ }\href {\doibase 10.1038/nature12483} {\bibfield
  {journal} {\bibinfo  {journal} {Nature}\ }\textbf {\bibinfo {volume} {501}},\
  \bibinfo {pages} {521} (\bibinfo {year} {2013})}\BibitemShut {NoStop}%
\bibitem [{\citenamefont {Carr}\ \emph {et~al.}(2009)\citenamefont {Carr},
  \citenamefont {DeMille}, \citenamefont {Krems},\ and\ \citenamefont
  {Ye}}]{Carr2009}%
  \BibitemOpen
  \bibfield  {author} {\bibinfo {author} {\bibfnamefont {L.~D.}\ \bibnamefont
  {Carr}}, \bibinfo {author} {\bibfnamefont {D.}~\bibnamefont {DeMille}},
  \bibinfo {author} {\bibfnamefont {R.~V.}\ \bibnamefont {Krems}}, \ and\
  \bibinfo {author} {\bibfnamefont {J.}~\bibnamefont {Ye}},\ }\href {\doibase
  10.1088/1367-2630/11/5/055049} {\bibfield  {journal} {\bibinfo  {journal}
  {New J. Phys.}\ }\textbf {\bibinfo {volume} {11}},\ \bibinfo {pages} {055049}
  (\bibinfo {year} {2009})}\BibitemShut {NoStop}%
\bibitem [{And()}]{Andreev2018}%
  \BibitemOpen
  \href {\doibase 10.1038/s41586-018-0599-8} {}\bibinfo {note} {V. Andreev et
  al. (ACME Collaboration), Nature \textbf{562}, 355 (2018).}\BibitemShut
  {Stop}%
\bibitem [{\citenamefont {DeMille}(2002)}]{Demille2002}%
  \BibitemOpen
  \bibfield  {author} {\bibinfo {author} {\bibfnamefont {D.}~\bibnamefont
  {DeMille}},\ }\href {\doibase 10.1103/PhysRevLett.88.067901} {\bibfield
  {journal} {\bibinfo  {journal} {Phys. Rev. Lett.}\ }\textbf {\bibinfo
  {volume} {88}},\ \bibinfo {pages} {067901} (\bibinfo {year}
  {2002})}\BibitemShut {NoStop}%
\bibitem [{\citenamefont {Yelin}\ \emph {et~al.}(2006)\citenamefont {Yelin},
  \citenamefont {Kirby},\ and\ \citenamefont {C\^ot\'e}}]{Yelin2006}%
  \BibitemOpen
  \bibfield  {author} {\bibinfo {author} {\bibfnamefont {S.~F.}\ \bibnamefont
  {Yelin}}, \bibinfo {author} {\bibfnamefont {K.}~\bibnamefont {Kirby}}, \ and\
  \bibinfo {author} {\bibfnamefont {R.}~\bibnamefont {C\^ot\'e}},\ }\href
  {\doibase 10.1103/PhysRevA.74.050301} {\bibfield  {journal} {\bibinfo
  {journal} {Phys. Rev. A}\ }\textbf {\bibinfo {volume} {74}},\ \bibinfo
  {pages} {050301} (\bibinfo {year} {2006})}\BibitemShut {NoStop}%
\bibitem [{\citenamefont {Park}\ \emph {et~al.}(2017)\citenamefont {Park},
  \citenamefont {Yan}, \citenamefont {Loh}, \citenamefont {Will},\ and\
  \citenamefont {Zwierlein}}]{Park2017}%
  \BibitemOpen
  \bibfield  {author} {\bibinfo {author} {\bibfnamefont {J.~W.}\ \bibnamefont
  {Park}}, \bibinfo {author} {\bibfnamefont {Z.~Z.}\ \bibnamefont {Yan}},
  \bibinfo {author} {\bibfnamefont {H.}~\bibnamefont {Loh}}, \bibinfo {author}
  {\bibfnamefont {S.~A.}\ \bibnamefont {Will}}, \ and\ \bibinfo {author}
  {\bibfnamefont {M.~W.}\ \bibnamefont {Zwierlein}},\ }\href {\doibase
  10.1126/science.aal5066} {\bibfield  {journal} {\bibinfo  {journal}
  {Science}\ }\textbf {\bibinfo {volume} {357}},\ \bibinfo {pages} {372}
  (\bibinfo {year} {2017})}\BibitemShut {NoStop}%
\bibitem [{\citenamefont {Krems}(2008)}]{Krems2008}%
  \BibitemOpen
  \bibfield  {author} {\bibinfo {author} {\bibfnamefont {R.~V.}\ \bibnamefont
  {Krems}},\ }\href {\doibase 10.1039/b802322k} {\bibfield  {journal} {\bibinfo
   {journal} {Phys. Chem. Chem. Phys.}\ }\textbf {\bibinfo {volume} {10}},\
  \bibinfo {pages} {4079} (\bibinfo {year} {2008})}\BibitemShut {NoStop}%
\bibitem [{\citenamefont {Qu{\'{e}}m{\'{e}}ner}\ and\ \citenamefont
  {Julienne}(2012)}]{Quemener2012}%
  \BibitemOpen
  \bibfield  {author} {\bibinfo {author} {\bibfnamefont {G.}~\bibnamefont
  {Qu{\'{e}}m{\'{e}}ner}}\ and\ \bibinfo {author} {\bibfnamefont {P.~S.}\
  \bibnamefont {Julienne}},\ }\href {\doibase 10.1021/cr300092g} {\bibfield
  {journal} {\bibinfo  {journal} {Chem. Rev.}\ }\textbf {\bibinfo {volume}
  {112}},\ \bibinfo {pages} {4949} (\bibinfo {year} {2012})}\BibitemShut
  {NoStop}%
\bibitem [{\citenamefont {Balakrishnan}(2016)}]{Balakrishnan2016}%
  \BibitemOpen
  \bibfield  {author} {\bibinfo {author} {\bibfnamefont {N.}~\bibnamefont
  {Balakrishnan}},\ }\href {\doibase 10.1063/1.4964096} {\bibfield  {journal}
  {\bibinfo  {journal} {J. Chem. Phys.}\ }\textbf {\bibinfo {volume} {145}},\
  \bibinfo {pages} {150901} (\bibinfo {year} {2016})}\BibitemShut {NoStop}%
\bibitem [{\citenamefont {Yang}\ \emph {et~al.}(2019)\citenamefont {Yang},
  \citenamefont {Zhang}, \citenamefont {Liu}, \citenamefont {Liu},
  \citenamefont {Nan}, \citenamefont {Zhao},\ and\ \citenamefont
  {Pan}}]{Yang2019}%
  \BibitemOpen
  \bibfield  {author} {\bibinfo {author} {\bibfnamefont {H.}~\bibnamefont
  {Yang}}, \bibinfo {author} {\bibfnamefont {D.-C.}\ \bibnamefont {Zhang}},
  \bibinfo {author} {\bibfnamefont {L.}~\bibnamefont {Liu}}, \bibinfo {author}
  {\bibfnamefont {Y.-X.}\ \bibnamefont {Liu}}, \bibinfo {author} {\bibfnamefont
  {J.}~\bibnamefont {Nan}}, \bibinfo {author} {\bibfnamefont {B.}~\bibnamefont
  {Zhao}}, \ and\ \bibinfo {author} {\bibfnamefont {J.-W.}\ \bibnamefont
  {Pan}},\ }\href {\doibase 10.1126/science.aau5322} {\bibfield  {journal}
  {\bibinfo  {journal} {Science}\ }\textbf {\bibinfo {volume} {363}},\ \bibinfo
  {pages} {261} (\bibinfo {year} {2019})}\BibitemShut {NoStop}%
\bibitem [{\citenamefont {Ni}\ \emph {et~al.}(2008)\citenamefont {Ni},
  \citenamefont {Ospelkaus}, \citenamefont {{De Miranda}}, \citenamefont
  {Pe'er}, \citenamefont {Neyenhuis}, \citenamefont {Zirbel}, \citenamefont
  {Kotochigova}, \citenamefont {Julienne}, \citenamefont {Jin},\ and\
  \citenamefont {Ye}}]{Ni2008}%
  \BibitemOpen
  \bibfield  {author} {\bibinfo {author} {\bibfnamefont {K.~K.}\ \bibnamefont
  {Ni}}, \bibinfo {author} {\bibfnamefont {S.}~\bibnamefont {Ospelkaus}},
  \bibinfo {author} {\bibfnamefont {M.~H.}\ \bibnamefont {{De Miranda}}},
  \bibinfo {author} {\bibfnamefont {A.}~\bibnamefont {Pe'er}}, \bibinfo
  {author} {\bibfnamefont {B.}~\bibnamefont {Neyenhuis}}, \bibinfo {author}
  {\bibfnamefont {J.~J.}\ \bibnamefont {Zirbel}}, \bibinfo {author}
  {\bibfnamefont {S.}~\bibnamefont {Kotochigova}}, \bibinfo {author}
  {\bibfnamefont {P.~S.}\ \bibnamefont {Julienne}}, \bibinfo {author}
  {\bibfnamefont {D.~S.}\ \bibnamefont {Jin}}, \ and\ \bibinfo {author}
  {\bibfnamefont {J.}~\bibnamefont {Ye}},\ }\href {\doibase
  10.1126/science.1163861} {\bibfield  {journal} {\bibinfo  {journal}
  {Science}\ }\textbf {\bibinfo {volume} {322}},\ \bibinfo {pages} {231}
  (\bibinfo {year} {2008})}\BibitemShut {NoStop}%
\bibitem [{\citenamefont {Danzl}\ \emph {et~al.}(2010)\citenamefont {Danzl},
  \citenamefont {Mark}, \citenamefont {Haller}, \citenamefont {Gustavsson},
  \citenamefont {Hart}, \citenamefont {Aldegunde}, \citenamefont {Hutson},\
  and\ \citenamefont {N{\"{a}}gerl}}]{Danzl2010}%
  \BibitemOpen
  \bibfield  {author} {\bibinfo {author} {\bibfnamefont {J.~G.}\ \bibnamefont
  {Danzl}}, \bibinfo {author} {\bibfnamefont {M.~J.}\ \bibnamefont {Mark}},
  \bibinfo {author} {\bibfnamefont {E.}~\bibnamefont {Haller}}, \bibinfo
  {author} {\bibfnamefont {M.}~\bibnamefont {Gustavsson}}, \bibinfo {author}
  {\bibfnamefont {R.}~\bibnamefont {Hart}}, \bibinfo {author} {\bibfnamefont
  {J.}~\bibnamefont {Aldegunde}}, \bibinfo {author} {\bibfnamefont {J.~M.}\
  \bibnamefont {Hutson}}, \ and\ \bibinfo {author} {\bibfnamefont {H.~C.}\
  \bibnamefont {N{\"{a}}gerl}},\ }\href {\doibase 10.1038/nphys1533} {\bibfield
   {journal} {\bibinfo  {journal} {Nat. Phys.}\ }\textbf {\bibinfo {volume}
  {6}},\ \bibinfo {pages} {265} (\bibinfo {year} {2010})}\BibitemShut {NoStop}%
\bibitem [{\citenamefont {Takekoshi}\ \emph {et~al.}(2014)\citenamefont
  {Takekoshi}, \citenamefont {Reichs{\"{o}}llner}, \citenamefont {Schindewolf},
  \citenamefont {Hutson}, \citenamefont {{Le Sueur}}, \citenamefont {Dulieu},
  \citenamefont {Ferlaino}, \citenamefont {Grimm},\ and\ \citenamefont
  {N{\"{a}}gerl}}]{Takekoshi2014}%
  \BibitemOpen
  \bibfield  {author} {\bibinfo {author} {\bibfnamefont {T.}~\bibnamefont
  {Takekoshi}}, \bibinfo {author} {\bibfnamefont {L.}~\bibnamefont
  {Reichs{\"{o}}llner}}, \bibinfo {author} {\bibfnamefont {A.}~\bibnamefont
  {Schindewolf}}, \bibinfo {author} {\bibfnamefont {J.~M.}\ \bibnamefont
  {Hutson}}, \bibinfo {author} {\bibfnamefont {C.~R.}\ \bibnamefont {{Le
  Sueur}}}, \bibinfo {author} {\bibfnamefont {O.}~\bibnamefont {Dulieu}},
  \bibinfo {author} {\bibfnamefont {F.}~\bibnamefont {Ferlaino}}, \bibinfo
  {author} {\bibfnamefont {R.}~\bibnamefont {Grimm}}, \ and\ \bibinfo {author}
  {\bibfnamefont {H.~C.}\ \bibnamefont {N{\"{a}}gerl}},\ }\href {\doibase
  10.1103/PhysRevLett.113.205301} {\bibfield  {journal} {\bibinfo  {journal}
  {Phys. Rev. Lett.}\ }\textbf {\bibinfo {volume} {113}},\ \bibinfo {pages}
  {205301} (\bibinfo {year} {2014})}\BibitemShut {NoStop}%
\bibitem [{\citenamefont {Molony}\ \emph {et~al.}(2014)\citenamefont {Molony},
  \citenamefont {Gregory}, \citenamefont {Ji}, \citenamefont {Lu},
  \citenamefont {K{\"{o}}ppinger}, \citenamefont {{Le Sueur}}, \citenamefont
  {Blackley}, \citenamefont {Hutson},\ and\ \citenamefont
  {Cornish}}]{Molony2014}%
  \BibitemOpen
  \bibfield  {author} {\bibinfo {author} {\bibfnamefont {P.~K.}\ \bibnamefont
  {Molony}}, \bibinfo {author} {\bibfnamefont {P.~D.}\ \bibnamefont {Gregory}},
  \bibinfo {author} {\bibfnamefont {Z.}~\bibnamefont {Ji}}, \bibinfo {author}
  {\bibfnamefont {B.}~\bibnamefont {Lu}}, \bibinfo {author} {\bibfnamefont
  {M.~P.}\ \bibnamefont {K{\"{o}}ppinger}}, \bibinfo {author} {\bibfnamefont
  {C.~R.}\ \bibnamefont {{Le Sueur}}}, \bibinfo {author} {\bibfnamefont
  {C.~L.}\ \bibnamefont {Blackley}}, \bibinfo {author} {\bibfnamefont {J.~M.}\
  \bibnamefont {Hutson}}, \ and\ \bibinfo {author} {\bibfnamefont {S.~L.}\
  \bibnamefont {Cornish}},\ }\href {\doibase 10.1103/PhysRevLett.113.255301}
  {\bibfield  {journal} {\bibinfo  {journal} {Phys. Rev. Lett.}\ }\textbf
  {\bibinfo {volume} {113}},\ \bibinfo {pages} {255301} (\bibinfo {year}
  {2014})}\BibitemShut {NoStop}%
\bibitem [{\citenamefont {Park}\ \emph
  {et~al.}(2015{\natexlab{a}})\citenamefont {Park}, \citenamefont {Will},\ and\
  \citenamefont {Zwierlein}}]{Park2015}%
  \BibitemOpen
  \bibfield  {author} {\bibinfo {author} {\bibfnamefont {J.~W.}\ \bibnamefont
  {Park}}, \bibinfo {author} {\bibfnamefont {S.~A.}\ \bibnamefont {Will}}, \
  and\ \bibinfo {author} {\bibfnamefont {M.~W.}\ \bibnamefont {Zwierlein}},\
  }\href {\doibase 10.1103/PhysRevLett.114.205302} {\bibfield  {journal}
  {\bibinfo  {journal} {Phys. Rev. Lett.}\ }\textbf {\bibinfo {volume} {114}},\
  \bibinfo {pages} {205302} (\bibinfo {year} {2015}{\natexlab{a}})}\BibitemShut
  {NoStop}%
\bibitem [{\citenamefont {Guo}\ \emph {et~al.}(2016)\citenamefont {Guo},
  \citenamefont {Zhu}, \citenamefont {Lu}, \citenamefont {Ye}, \citenamefont
  {Wang}, \citenamefont {Vexiau}, \citenamefont {Bouloufa-Maafa}, \citenamefont
  {Qu{\'{e}}m{\'{e}}ner}, \citenamefont {Dulieu},\ and\ \citenamefont
  {Wang}}]{Guo2016}%
  \BibitemOpen
  \bibfield  {author} {\bibinfo {author} {\bibfnamefont {M.}~\bibnamefont
  {Guo}}, \bibinfo {author} {\bibfnamefont {B.}~\bibnamefont {Zhu}}, \bibinfo
  {author} {\bibfnamefont {B.}~\bibnamefont {Lu}}, \bibinfo {author}
  {\bibfnamefont {X.}~\bibnamefont {Ye}}, \bibinfo {author} {\bibfnamefont
  {F.}~\bibnamefont {Wang}}, \bibinfo {author} {\bibfnamefont {R.}~\bibnamefont
  {Vexiau}}, \bibinfo {author} {\bibfnamefont {N.}~\bibnamefont
  {Bouloufa-Maafa}}, \bibinfo {author} {\bibfnamefont {G.}~\bibnamefont
  {Qu{\'{e}}m{\'{e}}ner}}, \bibinfo {author} {\bibfnamefont {O.}~\bibnamefont
  {Dulieu}}, \ and\ \bibinfo {author} {\bibfnamefont {D.}~\bibnamefont
  {Wang}},\ }\href {\doibase 10.1103/PhysRevLett.116.205303} {\bibfield
  {journal} {\bibinfo  {journal} {Phys. Rev. Lett.}\ }\textbf {\bibinfo
  {volume} {116}},\ \bibinfo {pages} {205303} (\bibinfo {year}
  {2016})}\BibitemShut {NoStop}%
\bibitem [{\citenamefont {Rvachov}\ \emph {et~al.}(2017)\citenamefont
  {Rvachov}, \citenamefont {Son}, \citenamefont {Sommer}, \citenamefont
  {Ebadi}, \citenamefont {Park}, \citenamefont {Zwierlein}, \citenamefont
  {Ketterle},\ and\ \citenamefont {Jamison}}]{Rvachov2017}%
  \BibitemOpen
  \bibfield  {author} {\bibinfo {author} {\bibfnamefont {T.~M.}\ \bibnamefont
  {Rvachov}}, \bibinfo {author} {\bibfnamefont {H.}~\bibnamefont {Son}},
  \bibinfo {author} {\bibfnamefont {A.~T.}\ \bibnamefont {Sommer}}, \bibinfo
  {author} {\bibfnamefont {S.}~\bibnamefont {Ebadi}}, \bibinfo {author}
  {\bibfnamefont {J.~J.}\ \bibnamefont {Park}}, \bibinfo {author}
  {\bibfnamefont {M.~W.}\ \bibnamefont {Zwierlein}}, \bibinfo {author}
  {\bibfnamefont {W.}~\bibnamefont {Ketterle}}, \ and\ \bibinfo {author}
  {\bibfnamefont {A.~O.}\ \bibnamefont {Jamison}},\ }\href {\doibase
  10.1103/PhysRevLett.119.143001} {\bibfield  {journal} {\bibinfo  {journal}
  {Phys. Rev. Lett.}\ }\textbf {\bibinfo {volume} {119}},\ \bibinfo {pages}
  {143001} (\bibinfo {year} {2017})}\BibitemShut {NoStop}%
\bibitem [{\citenamefont {See\ss{}elberg}\ \emph {et~al.}(2018)\citenamefont
  {See\ss{}elberg}, \citenamefont {Buchheim}, \citenamefont {Lu}, \citenamefont
  {Schneider}, \citenamefont {Luo}, \citenamefont {Tiemann}, \citenamefont
  {Bloch},\ and\ \citenamefont {Gohle}}]{Seesselberg2018}%
  \BibitemOpen
  \bibfield  {author} {\bibinfo {author} {\bibfnamefont {F.}~\bibnamefont
  {See\ss{}elberg}}, \bibinfo {author} {\bibfnamefont {N.}~\bibnamefont
  {Buchheim}}, \bibinfo {author} {\bibfnamefont {Z.-K.}\ \bibnamefont {Lu}},
  \bibinfo {author} {\bibfnamefont {T.}~\bibnamefont {Schneider}}, \bibinfo
  {author} {\bibfnamefont {X.-Y.}\ \bibnamefont {Luo}}, \bibinfo {author}
  {\bibfnamefont {E.}~\bibnamefont {Tiemann}}, \bibinfo {author} {\bibfnamefont
  {I.}~\bibnamefont {Bloch}}, \ and\ \bibinfo {author} {\bibfnamefont
  {C.}~\bibnamefont {Gohle}},\ }\href {\doibase 10.1103/PhysRevA.97.013405}
  {\bibfield  {journal} {\bibinfo  {journal} {Phys. Rev. A}\ }\textbf {\bibinfo
  {volume} {97}},\ \bibinfo {pages} {013405} (\bibinfo {year}
  {2018})}\BibitemShut {NoStop}%
\bibitem [{\citenamefont {Micheli}\ \emph {et~al.}(2007)\citenamefont
  {Micheli}, \citenamefont {Pupillo}, \citenamefont {B{\"{u}}chler},\ and\
  \citenamefont {Zoller}}]{Micheli2007}%
  \BibitemOpen
  \bibfield  {author} {\bibinfo {author} {\bibfnamefont {A.}~\bibnamefont
  {Micheli}}, \bibinfo {author} {\bibfnamefont {G.}~\bibnamefont {Pupillo}},
  \bibinfo {author} {\bibfnamefont {H.~P.}\ \bibnamefont {B{\"{u}}chler}}, \
  and\ \bibinfo {author} {\bibfnamefont {P.}~\bibnamefont {Zoller}},\ }\href
  {\doibase 10.1103/PhysRevA.76.043604} {\bibfield  {journal} {\bibinfo
  {journal} {Phys. Rev. A}\ }\textbf {\bibinfo {volume} {76}},\ \bibinfo
  {pages} {043604} (\bibinfo {year} {2007})}\BibitemShut {NoStop}%
\bibitem [{\citenamefont {Bruun}\ and\ \citenamefont
  {Taylor}(2008)}]{Bruun2008}%
  \BibitemOpen
  \bibfield  {author} {\bibinfo {author} {\bibfnamefont {G.~M.}\ \bibnamefont
  {Bruun}}\ and\ \bibinfo {author} {\bibfnamefont {E.}~\bibnamefont {Taylor}},\
  }\href {\doibase 10.1103/PhysRevLett.101.245301} {\bibfield  {journal}
  {\bibinfo  {journal} {Phys. Rev. Lett.}\ }\textbf {\bibinfo {volume} {101}},\
  \bibinfo {pages} {245301} (\bibinfo {year} {2008})}\BibitemShut {NoStop}%
\bibitem [{\citenamefont {Cooper}\ and\ \citenamefont
  {Shlyapnikov}(2009)}]{Cooper2009}%
  \BibitemOpen
  \bibfield  {author} {\bibinfo {author} {\bibfnamefont {N.~R.}\ \bibnamefont
  {Cooper}}\ and\ \bibinfo {author} {\bibfnamefont {G.~V.}\ \bibnamefont
  {Shlyapnikov}},\ }\href {\doibase 10.1103/PhysRevLett.103.155302} {\bibfield
  {journal} {\bibinfo  {journal} {Phys. Rev. Lett.}\ }\textbf {\bibinfo
  {volume} {103}},\ \bibinfo {pages} {155302} (\bibinfo {year}
  {2009})}\BibitemShut {NoStop}%
\bibitem [{\citenamefont {Levinsen}\ \emph {et~al.}(2011)\citenamefont
  {Levinsen}, \citenamefont {Cooper},\ and\ \citenamefont
  {Shlyapnikov}}]{Levinsen2011}%
  \BibitemOpen
  \bibfield  {author} {\bibinfo {author} {\bibfnamefont {J.}~\bibnamefont
  {Levinsen}}, \bibinfo {author} {\bibfnamefont {N.~R.}\ \bibnamefont
  {Cooper}}, \ and\ \bibinfo {author} {\bibfnamefont {G.~V.}\ \bibnamefont
  {Shlyapnikov}},\ }\href {\doibase 10.1103/PhysRevA.84.013603} {\bibfield
  {journal} {\bibinfo  {journal} {Phys. Rev. A}\ }\textbf {\bibinfo {volume}
  {84}},\ \bibinfo {pages} {013603} (\bibinfo {year} {2011})}\BibitemShut
  {NoStop}%
\bibitem [{\citenamefont {Wall}\ and\ \citenamefont {Carr}(2010)}]{Wall2010}%
  \BibitemOpen
  \bibfield  {author} {\bibinfo {author} {\bibfnamefont {M.~L.}\ \bibnamefont
  {Wall}}\ and\ \bibinfo {author} {\bibfnamefont {L.~D.}\ \bibnamefont
  {Carr}},\ }\href {\doibase 10.1103/PhysRevA.82.013611} {\bibfield  {journal}
  {\bibinfo  {journal} {Phys. Rev. A}\ }\textbf {\bibinfo {volume} {82}},\
  \bibinfo {pages} {013611} (\bibinfo {year} {2010})}\BibitemShut {NoStop}%
\bibitem [{\citenamefont {Gorshkov}\ \emph {et~al.}(2011)\citenamefont
  {Gorshkov}, \citenamefont {Manmana}, \citenamefont {Chen}, \citenamefont
  {Ye}, \citenamefont {Demler}, \citenamefont {Lukin},\ and\ \citenamefont
  {Rey}}]{Gorshkov2011}%
  \BibitemOpen
  \bibfield  {author} {\bibinfo {author} {\bibfnamefont {A.~V.}\ \bibnamefont
  {Gorshkov}}, \bibinfo {author} {\bibfnamefont {S.~R.}\ \bibnamefont
  {Manmana}}, \bibinfo {author} {\bibfnamefont {G.}~\bibnamefont {Chen}},
  \bibinfo {author} {\bibfnamefont {J.}~\bibnamefont {Ye}}, \bibinfo {author}
  {\bibfnamefont {E.}~\bibnamefont {Demler}}, \bibinfo {author} {\bibfnamefont
  {M.~D.}\ \bibnamefont {Lukin}}, \ and\ \bibinfo {author} {\bibfnamefont
  {A.~M.}\ \bibnamefont {Rey}},\ }\href {\doibase
  10.1103/PhysRevLett.107.115301} {\bibfield  {journal} {\bibinfo  {journal}
  {Phys. Rev. Lett.}\ }\textbf {\bibinfo {volume} {107}},\ \bibinfo {pages}
  {115301} (\bibinfo {year} {2011})}\BibitemShut {NoStop}%
\bibitem [{\citenamefont {DeMille}\ \emph {et~al.}(2004)\citenamefont
  {DeMille}, \citenamefont {Glenn},\ and\ \citenamefont
  {Petricka}}]{DeMille2004}%
  \BibitemOpen
  \bibfield  {author} {\bibinfo {author} {\bibfnamefont {D.}~\bibnamefont
  {DeMille}}, \bibinfo {author} {\bibfnamefont {D.~R.}\ \bibnamefont {Glenn}},
  \ and\ \bibinfo {author} {\bibfnamefont {J.}~\bibnamefont {Petricka}},\
  }\href {\doibase 10.1140/epjd/e2004-00163-6} {\bibfield  {journal} {\bibinfo
  {journal} {Eur. Phys. J. D}\ }\textbf {\bibinfo {volume} {31}},\ \bibinfo
  {pages} {375} (\bibinfo {year} {2004})}\BibitemShut {NoStop}%
\bibitem [{\citenamefont {Dunseith}\ \emph {et~al.}(2015)\citenamefont
  {Dunseith}, \citenamefont {Truppe}, \citenamefont {Hendricks}, \citenamefont
  {Sauer}, \citenamefont {Hinds},\ and\ \citenamefont
  {Tarbutt}}]{Dunseith2015}%
  \BibitemOpen
  \bibfield  {author} {\bibinfo {author} {\bibfnamefont {D.~P.}\ \bibnamefont
  {Dunseith}}, \bibinfo {author} {\bibfnamefont {S.}~\bibnamefont {Truppe}},
  \bibinfo {author} {\bibfnamefont {R.~J.}\ \bibnamefont {Hendricks}}, \bibinfo
  {author} {\bibfnamefont {B.~E.}\ \bibnamefont {Sauer}}, \bibinfo {author}
  {\bibfnamefont {E.~A.}\ \bibnamefont {Hinds}}, \ and\ \bibinfo {author}
  {\bibfnamefont {M.~R.}\ \bibnamefont {Tarbutt}},\ }\href {\doibase
  10.1088/0953-4075/48/4/045001} {\bibfield  {journal} {\bibinfo  {journal} {J.
  Phys. B}\ }\textbf {\bibinfo {volume} {48}},\ \bibinfo {pages} {045001}
  (\bibinfo {year} {2015})}\BibitemShut {NoStop}%
\bibitem [{\citenamefont {Wright}\ \emph {et~al.}(2019)\citenamefont {Wright},
  \citenamefont {Wall},\ and\ \citenamefont {Tarbutt}}]{Wright2019}%
  \BibitemOpen
  \bibfield  {author} {\bibinfo {author} {\bibfnamefont {S.~C.}\ \bibnamefont
  {Wright}}, \bibinfo {author} {\bibfnamefont {T.~E.}\ \bibnamefont {Wall}}, \
  and\ \bibinfo {author} {\bibfnamefont {M.~R.}\ \bibnamefont {Tarbutt}},\
  }\href {\doibase 10.1103/PhysRevResearch.1.033035} {\bibfield  {journal}
  {\bibinfo  {journal} {Phys. Rev. Research}\ }\textbf {\bibinfo {volume}
  {1}},\ \bibinfo {pages} {033035} (\bibinfo {year} {2019})}\BibitemShut
  {NoStop}%
\bibitem [{\citenamefont {Gonz\'alez-Mart\'{\i}nez}\ \emph
  {et~al.}(2017)\citenamefont {Gonz\'alez-Mart\'{\i}nez}, \citenamefont
  {Bohn},\ and\ \citenamefont {Qu\'em\'ener}}]{Gonzalez-Martinez2017}%
  \BibitemOpen
  \bibfield  {author} {\bibinfo {author} {\bibfnamefont {M.~L.}\ \bibnamefont
  {Gonz\'alez-Mart\'{\i}nez}}, \bibinfo {author} {\bibfnamefont {J.~L.}\
  \bibnamefont {Bohn}}, \ and\ \bibinfo {author} {\bibfnamefont
  {G.}~\bibnamefont {Qu\'em\'ener}},\ }\href {\doibase
  10.1103/PhysRevA.96.032718} {\bibfield  {journal} {\bibinfo  {journal} {Phys.
  Rev. A}\ }\textbf {\bibinfo {volume} {96}},\ \bibinfo {pages} {032718}
  (\bibinfo {year} {2017})}\BibitemShut {NoStop}%
\bibitem [{\citenamefont {Gorshkov}\ \emph {et~al.}(2008)\citenamefont
  {Gorshkov}, \citenamefont {Rabl}, \citenamefont {Pupillo}, \citenamefont
  {Micheli}, \citenamefont {Zoller}, \citenamefont {Lukin},\ and\ \citenamefont
  {B{\"{u}}chler}}]{Gorshkov2008}%
  \BibitemOpen
  \bibfield  {author} {\bibinfo {author} {\bibfnamefont {A.~V.}\ \bibnamefont
  {Gorshkov}}, \bibinfo {author} {\bibfnamefont {P.}~\bibnamefont {Rabl}},
  \bibinfo {author} {\bibfnamefont {G.}~\bibnamefont {Pupillo}}, \bibinfo
  {author} {\bibfnamefont {A.}~\bibnamefont {Micheli}}, \bibinfo {author}
  {\bibfnamefont {P.}~\bibnamefont {Zoller}}, \bibinfo {author} {\bibfnamefont
  {M.~D.}\ \bibnamefont {Lukin}}, \ and\ \bibinfo {author} {\bibfnamefont
  {H.~P.}\ \bibnamefont {B{\"{u}}chler}},\ }\href {\doibase
  10.1103/PhysRevLett.101.073201} {\bibfield  {journal} {\bibinfo  {journal}
  {Phys. Rev. Lett.}\ }\textbf {\bibinfo {volume} {101}},\ \bibinfo {pages}
  {073201} (\bibinfo {year} {2008})}\BibitemShut {NoStop}%
\bibitem [{\citenamefont {Karman}\ and\ \citenamefont
  {Hutson}(2018)}]{Karman2018a}%
  \BibitemOpen
  \bibfield  {author} {\bibinfo {author} {\bibfnamefont {T.}~\bibnamefont
  {Karman}}\ and\ \bibinfo {author} {\bibfnamefont {J.~M.}\ \bibnamefont
  {Hutson}},\ }\href {\doibase 10.1103/PhysRevLett.121.163401} {\bibfield
  {journal} {\bibinfo  {journal} {Phys. Rev. Lett.}\ }\textbf {\bibinfo
  {volume} {121}},\ \bibinfo {pages} {163401} (\bibinfo {year}
  {2018})}\BibitemShut {NoStop}%
\bibitem [{\citenamefont {Lassabli\`ere}\ and\ \citenamefont
  {Qu\'em\'ener}(2018)}]{lassabliere:18}%
  \BibitemOpen
  \bibfield  {author} {\bibinfo {author} {\bibfnamefont {L.}~\bibnamefont
  {Lassabli\`ere}}\ and\ \bibinfo {author} {\bibfnamefont {G.}~\bibnamefont
  {Qu\'em\'ener}},\ }\href {\doibase 10.1103/PhysRevLett.121.163402} {\bibfield
   {journal} {\bibinfo  {journal} {Phys. Rev. Lett.}\ }\textbf {\bibinfo
  {volume} {121}},\ \bibinfo {pages} {163402} (\bibinfo {year}
  {2018})}\BibitemShut {NoStop}%
\bibitem [{\citenamefont {Ospelkaus}\ \emph {et~al.}(2010)\citenamefont
  {Ospelkaus}, \citenamefont {Ni}, \citenamefont {Wang}, \citenamefont
  {de~Miranda}, \citenamefont {Neyenhuis}, \citenamefont {Qu{\'e}m{\'e}ner},
  \citenamefont {Julienne}, \citenamefont {Bohn}, \citenamefont {Jin},\ and\
  \citenamefont {Ye}}]{Ospelkaus2010}%
  \BibitemOpen
  \bibfield  {author} {\bibinfo {author} {\bibfnamefont {S.}~\bibnamefont
  {Ospelkaus}}, \bibinfo {author} {\bibfnamefont {K.-K.}\ \bibnamefont {Ni}},
  \bibinfo {author} {\bibfnamefont {D.}~\bibnamefont {Wang}}, \bibinfo {author}
  {\bibfnamefont {M.~H.~G.}\ \bibnamefont {de~Miranda}}, \bibinfo {author}
  {\bibfnamefont {B.}~\bibnamefont {Neyenhuis}}, \bibinfo {author}
  {\bibfnamefont {G.}~\bibnamefont {Qu{\'e}m{\'e}ner}}, \bibinfo {author}
  {\bibfnamefont {P.~S.}\ \bibnamefont {Julienne}}, \bibinfo {author}
  {\bibfnamefont {J.~L.}\ \bibnamefont {Bohn}}, \bibinfo {author}
  {\bibfnamefont {D.~S.}\ \bibnamefont {Jin}}, \ and\ \bibinfo {author}
  {\bibfnamefont {J.}~\bibnamefont {Ye}},\ }\href {\doibase
  10.1126/science.1184121} {\bibfield  {journal} {\bibinfo  {journal}
  {Science}\ }\textbf {\bibinfo {volume} {327}},\ \bibinfo {pages} {853}
  (\bibinfo {year} {2010})}\BibitemShut {NoStop}%
\bibitem [{\citenamefont {Ye}\ \emph {et~al.}(2018)\citenamefont {Ye},
  \citenamefont {Guo}, \citenamefont {Gonz{\'{a}}lez-Mart{\'{i}}nez},
  \citenamefont {Qu{\'{e}}m{\'{e}}ner},\ and\ \citenamefont {Wang}}]{Ye2018}%
  \BibitemOpen
  \bibfield  {author} {\bibinfo {author} {\bibfnamefont {X.}~\bibnamefont
  {Ye}}, \bibinfo {author} {\bibfnamefont {M.}~\bibnamefont {Guo}}, \bibinfo
  {author} {\bibfnamefont {M.~L.}\ \bibnamefont
  {Gonz{\'{a}}lez-Mart{\'{i}}nez}}, \bibinfo {author} {\bibfnamefont
  {G.}~\bibnamefont {Qu{\'{e}}m{\'{e}}ner}}, \ and\ \bibinfo {author}
  {\bibfnamefont {D.}~\bibnamefont {Wang}},\ }\href {\doibase
  10.1126/sciadv.aaq0083} {\bibfield  {journal} {\bibinfo  {journal} {Sci.
  Adv.}\ }\textbf {\bibinfo {volume} {4}},\ \bibinfo {pages} {eaaq0083}
  (\bibinfo {year} {2018})}\BibitemShut {NoStop}%
\bibitem [{\citenamefont {Will}\ \emph {et~al.}(2016)\citenamefont {Will},
  \citenamefont {Park}, \citenamefont {Yan}, \citenamefont {Loh},\ and\
  \citenamefont {Zwierlein}}]{Will2016}%
  \BibitemOpen
  \bibfield  {author} {\bibinfo {author} {\bibfnamefont {S.~A.}\ \bibnamefont
  {Will}}, \bibinfo {author} {\bibfnamefont {J.~W.}\ \bibnamefont {Park}},
  \bibinfo {author} {\bibfnamefont {Z.~Z.}\ \bibnamefont {Yan}}, \bibinfo
  {author} {\bibfnamefont {H.}~\bibnamefont {Loh}}, \ and\ \bibinfo {author}
  {\bibfnamefont {M.~W.}\ \bibnamefont {Zwierlein}},\ }\href {\doibase
  10.1103/PhysRevLett.116.225306} {\bibfield  {journal} {\bibinfo  {journal}
  {Phys. Rev. Lett.}\ }\textbf {\bibinfo {volume} {116}},\ \bibinfo {pages}
  {225306} (\bibinfo {year} {2016})}\BibitemShut {NoStop}%
\bibitem [{\citenamefont {Guo}\ \emph {et~al.}(2018)\citenamefont {Guo},
  \citenamefont {Ye}, \citenamefont {He}, \citenamefont
  {Gonz{\'{a}}lez-Mart{\'{i}}nez}, \citenamefont {Vexiau}, \citenamefont
  {Qu{\'{e}}m{\'{e}}ner},\ and\ \citenamefont {Wang}}]{Guo2018}%
  \BibitemOpen
  \bibfield  {author} {\bibinfo {author} {\bibfnamefont {M.}~\bibnamefont
  {Guo}}, \bibinfo {author} {\bibfnamefont {X.}~\bibnamefont {Ye}}, \bibinfo
  {author} {\bibfnamefont {J.}~\bibnamefont {He}}, \bibinfo {author}
  {\bibfnamefont {M.~L.}\ \bibnamefont {Gonz{\'{a}}lez-Mart{\'{i}}nez}},
  \bibinfo {author} {\bibfnamefont {R.}~\bibnamefont {Vexiau}}, \bibinfo
  {author} {\bibfnamefont {G.}~\bibnamefont {Qu{\'{e}}m{\'{e}}ner}}, \ and\
  \bibinfo {author} {\bibfnamefont {D.}~\bibnamefont {Wang}},\ }\href {\doibase
  10.1103/PhysRevX.8.041044} {\bibfield  {journal} {\bibinfo  {journal} {Phys.
  Rev. X}\ }\textbf {\bibinfo {volume} {8}},\ \bibinfo {pages} {041044}
  (\bibinfo {year} {2018})}\BibitemShut {NoStop}%
\bibitem [{\citenamefont {Gregory}\ \emph {et~al.}(2019)\citenamefont
  {Gregory}, \citenamefont {Frye}, \citenamefont {Blackmore}, \citenamefont
  {Bridge}, \citenamefont {Sawant}, \citenamefont {Hutson},\ and\ \citenamefont
  {Cornish}}]{Gregory2019}%
  \BibitemOpen
  \bibfield  {author} {\bibinfo {author} {\bibfnamefont {P.~D.}\ \bibnamefont
  {Gregory}}, \bibinfo {author} {\bibfnamefont {M.~D.}\ \bibnamefont {Frye}},
  \bibinfo {author} {\bibfnamefont {J.~A.}\ \bibnamefont {Blackmore}}, \bibinfo
  {author} {\bibfnamefont {E.~M.}\ \bibnamefont {Bridge}}, \bibinfo {author}
  {\bibfnamefont {R.}~\bibnamefont {Sawant}}, \bibinfo {author} {\bibfnamefont
  {J.~M.}\ \bibnamefont {Hutson}}, \ and\ \bibinfo {author} {\bibfnamefont
  {S.~L.}\ \bibnamefont {Cornish}},\ }\href {\doibase
  10.1038/s41467-019-11033-y} {\bibfield  {journal} {\bibinfo  {journal} {Nat.
  Commun.}\ }\textbf {\bibinfo {volume} {10}},\ \bibinfo {pages} {3104}
  (\bibinfo {year} {2019})}\BibitemShut {NoStop}%
\bibitem [{\citenamefont {Christianen}\ \emph {et~al.}(2019)\citenamefont
  {Christianen}, \citenamefont {Zwierlein}, \citenamefont {Groenenboom},\ and\
  \citenamefont {Karman}}]{Christianen2019}%
  \BibitemOpen
  \bibfield  {author} {\bibinfo {author} {\bibfnamefont {A.}~\bibnamefont
  {Christianen}}, \bibinfo {author} {\bibfnamefont {M.~W.}\ \bibnamefont
  {Zwierlein}}, \bibinfo {author} {\bibfnamefont {G.~C.}\ \bibnamefont
  {Groenenboom}}, \ and\ \bibinfo {author} {\bibfnamefont {T.}~\bibnamefont
  {Karman}},\ }\href {\doibase 10.1103/PhysRevLett.123.123402} {\bibfield
  {journal} {\bibinfo  {journal} {Phys. Rev. Lett.}\ }\textbf {\bibinfo
  {volume} {123}},\ \bibinfo {pages} {123402} (\bibinfo {year}
  {2019})}\BibitemShut {NoStop}%
\bibitem [{\citenamefont {Gaunt}\ \emph {et~al.}(2013)\citenamefont {Gaunt},
  \citenamefont {Schmidutz}, \citenamefont {Gotlibovych}, \citenamefont
  {Smith},\ and\ \citenamefont {Hadzibabic}}]{gaunt:13}%
  \BibitemOpen
  \bibfield  {author} {\bibinfo {author} {\bibfnamefont {A.~L.}\ \bibnamefont
  {Gaunt}}, \bibinfo {author} {\bibfnamefont {T.~F.}\ \bibnamefont
  {Schmidutz}}, \bibinfo {author} {\bibfnamefont {I.}~\bibnamefont
  {Gotlibovych}}, \bibinfo {author} {\bibfnamefont {R.~P.}\ \bibnamefont
  {Smith}}, \ and\ \bibinfo {author} {\bibfnamefont {Z.}~\bibnamefont
  {Hadzibabic}},\ }\href {\doibase 10.1103/PhysRevLett.110.200406} {\bibfield
  {journal} {\bibinfo  {journal} {Phys. Rev. Lett.}\ }\textbf {\bibinfo
  {volume} {110}},\ \bibinfo {pages} {200406} (\bibinfo {year}
  {2013})}\BibitemShut {NoStop}%
\bibitem [{\citenamefont {Mukherjee}\ \emph {et~al.}(2017)\citenamefont
  {Mukherjee}, \citenamefont {Yan}, \citenamefont {Patel}, \citenamefont
  {Hadzibabic}, \citenamefont {Yefsah}, \citenamefont {Struck},\ and\
  \citenamefont {Zwierlein}}]{mukherjee:17}%
  \BibitemOpen
  \bibfield  {author} {\bibinfo {author} {\bibfnamefont {B.}~\bibnamefont
  {Mukherjee}}, \bibinfo {author} {\bibfnamefont {Z.}~\bibnamefont {Yan}},
  \bibinfo {author} {\bibfnamefont {P.~B.}\ \bibnamefont {Patel}}, \bibinfo
  {author} {\bibfnamefont {Z.}~\bibnamefont {Hadzibabic}}, \bibinfo {author}
  {\bibfnamefont {T.}~\bibnamefont {Yefsah}}, \bibinfo {author} {\bibfnamefont
  {J.}~\bibnamefont {Struck}}, \ and\ \bibinfo {author} {\bibfnamefont {M.~W.}\
  \bibnamefont {Zwierlein}},\ }\href {\doibase 10.1103/PhysRevLett.118.123401}
  {\bibfield  {journal} {\bibinfo  {journal} {Phys. Rev. Lett.}\ }\textbf
  {\bibinfo {volume} {118}},\ \bibinfo {pages} {123401} (\bibinfo {year}
  {2017})}\BibitemShut {NoStop}%
\bibitem [{\citenamefont {Julienne}(1996)}]{Julienne1996}%
  \BibitemOpen
  \bibfield  {author} {\bibinfo {author} {\bibfnamefont {P.~S.}\ \bibnamefont
  {Julienne}},\ }\href {\doibase 10.6028/jres.101.050} {\bibfield  {journal}
  {\bibinfo  {journal} {Journal of Research of the National Institute of
  Standards and Technology}\ }\textbf {\bibinfo {volume} {101}},\ \bibinfo
  {pages} {487} (\bibinfo {year} {1996})}\BibitemShut {NoStop}%
\bibitem [{\citenamefont {Boisseau}\ \emph {et~al.}(2000)\citenamefont
  {Boisseau}, \citenamefont {Audouard}, \citenamefont {Vigu\'e},\ and\
  \citenamefont {Julienne}}]{boisseau2000reflection}%
  \BibitemOpen
  \bibfield  {author} {\bibinfo {author} {\bibfnamefont {C.}~\bibnamefont
  {Boisseau}}, \bibinfo {author} {\bibfnamefont {E.}~\bibnamefont {Audouard}},
  \bibinfo {author} {\bibfnamefont {J.}~\bibnamefont {Vigu\'e}}, \ and\
  \bibinfo {author} {\bibfnamefont {P.~S.}\ \bibnamefont {Julienne}},\ }\href
  {\doibase 10.1103/PhysRevA.62.052705} {\bibfield  {journal} {\bibinfo
  {journal} {Phys. Rev. A}\ }\textbf {\bibinfo {volume} {62}},\ \bibinfo
  {pages} {052705} (\bibinfo {year} {2000})}\BibitemShut {NoStop}%
\bibitem [{\citenamefont {Park}\ \emph
  {et~al.}(2015{\natexlab{b}})\citenamefont {Park}, \citenamefont {Will},\ and\
  \citenamefont {Zwierlein}}]{Park2015a}%
  \BibitemOpen
  \bibfield  {author} {\bibinfo {author} {\bibfnamefont {J.~W.}\ \bibnamefont
  {Park}}, \bibinfo {author} {\bibfnamefont {S.~A.}\ \bibnamefont {Will}}, \
  and\ \bibinfo {author} {\bibfnamefont {M.~W.}\ \bibnamefont {Zwierlein}},\
  }\href {\doibase 10.1088/1367-2630/17/7/075016} {\bibfield  {journal}
  {\bibinfo  {journal} {New Journal of Physics}\ }\textbf {\bibinfo {volume}
  {17}},\ \bibinfo {pages} {075016} (\bibinfo {year}
  {2015}{\natexlab{b}})}\BibitemShut {NoStop}%
\bibitem [{Sup()}]{Supplement}%
  \BibitemOpen
  \href@noop {} {}\bibinfo {note} {See supplement}\BibitemShut {NoStop}%
\bibitem [{\citenamefont {Idziaszek}\ and\ \citenamefont
  {Julienne}(2010)}]{Idziaszek2010}%
  \BibitemOpen
  \bibfield  {author} {\bibinfo {author} {\bibfnamefont {Z.}~\bibnamefont
  {Idziaszek}}\ and\ \bibinfo {author} {\bibfnamefont {P.~S.}\ \bibnamefont
  {Julienne}},\ }\href {\doibase 10.1103/PhysRevLett.104.113202} {\bibfield
  {journal} {\bibinfo  {journal} {Phys. Rev. Lett.}\ }\textbf {\bibinfo
  {volume} {104}},\ \bibinfo {pages} {113202} (\bibinfo {year}
  {2010})}\BibitemShut {NoStop}%
\bibitem [{\citenamefont {Ni}\ \emph {et~al.}(2010)\citenamefont {Ni},
  \citenamefont {Ospelkaus}, \citenamefont {Wang}, \citenamefont
  {Qu{\'{e}}m{\'{e}}ner}, \citenamefont {Neyenhuis}, \citenamefont {{De
  Miranda}}, \citenamefont {Bohn}, \citenamefont {Ye},\ and\ \citenamefont
  {Jin}}]{Ni2010}%
  \BibitemOpen
  \bibfield  {author} {\bibinfo {author} {\bibfnamefont {K.~K.}\ \bibnamefont
  {Ni}}, \bibinfo {author} {\bibfnamefont {S.}~\bibnamefont {Ospelkaus}},
  \bibinfo {author} {\bibfnamefont {D.}~\bibnamefont {Wang}}, \bibinfo {author}
  {\bibfnamefont {G.}~\bibnamefont {Qu{\'{e}}m{\'{e}}ner}}, \bibinfo {author}
  {\bibfnamefont {B.}~\bibnamefont {Neyenhuis}}, \bibinfo {author}
  {\bibfnamefont {M.~H.}\ \bibnamefont {{De Miranda}}}, \bibinfo {author}
  {\bibfnamefont {J.~L.}\ \bibnamefont {Bohn}}, \bibinfo {author}
  {\bibfnamefont {J.}~\bibnamefont {Ye}}, \ and\ \bibinfo {author}
  {\bibfnamefont {D.~S.}\ \bibnamefont {Jin}},\ }\href {\doibase
  10.1038/nature08953} {\bibfield  {journal} {\bibinfo  {journal} {Nature}\
  }\textbf {\bibinfo {volume} {464}},\ \bibinfo {pages} {1324} (\bibinfo {year}
  {2010})}\BibitemShut {NoStop}%
\bibitem [{\citenamefont {Burnett}\ \emph {et~al.}(1996)\citenamefont
  {Burnett}, \citenamefont {Julienne},\ and\ \citenamefont
  {Suominen}}]{burnett1996laser}%
  \BibitemOpen
  \bibfield  {author} {\bibinfo {author} {\bibfnamefont {K.}~\bibnamefont
  {Burnett}}, \bibinfo {author} {\bibfnamefont {P.~S.}\ \bibnamefont
  {Julienne}}, \ and\ \bibinfo {author} {\bibfnamefont {K.-A.}\ \bibnamefont
  {Suominen}},\ }\href {\doibase 10.1103/PhysRevLett.77.1416} {\bibfield
  {journal} {\bibinfo  {journal} {Phys. Rev. Lett.}\ }\textbf {\bibinfo
  {volume} {77}},\ \bibinfo {pages} {1416} (\bibinfo {year}
  {1996})}\BibitemShut {NoStop}%
\bibitem [{\citenamefont {Janssen}\ \emph {et~al.}(2013)\citenamefont
  {Janssen}, \citenamefont {van~der Avoird},\ and\ \citenamefont
  {Groenenboom}}]{janssen:13}%
  \BibitemOpen
  \bibfield  {author} {\bibinfo {author} {\bibfnamefont {L.~M.~C.}\
  \bibnamefont {Janssen}}, \bibinfo {author} {\bibfnamefont {A.}~\bibnamefont
  {van~der Avoird}}, \ and\ \bibinfo {author} {\bibfnamefont {G.~C.}\
  \bibnamefont {Groenenboom}},\ }\href {\doibase
  10.1103/PhysRevLett.110.063201} {\bibfield  {journal} {\bibinfo  {journal}
  {Phys. Rev. Lett.}\ }\textbf {\bibinfo {volume} {110}},\ \bibinfo {pages}
  {063201} (\bibinfo {year} {2013})}\BibitemShut {NoStop}%
\bibitem [{\citenamefont {Aldegunde}\ and\ \citenamefont
  {Hutson}(2017)}]{aldegunde:17}%
  \BibitemOpen
  \bibfield  {author} {\bibinfo {author} {\bibfnamefont {J.}~\bibnamefont
  {Aldegunde}}\ and\ \bibinfo {author} {\bibfnamefont {J.~M.}\ \bibnamefont
  {Hutson}},\ }\href {\doibase 10.1103/PhysRevA.96.042506} {\bibfield
  {journal} {\bibinfo  {journal} {Phys. Rev. A}\ }\textbf {\bibinfo {volume}
  {96}},\ \bibinfo {pages} {042506} (\bibinfo {year} {2017})}\BibitemShut
  {NoStop}%
\bibitem [{\citenamefont {Aldegunde}\ \emph {et~al.}(2008)\citenamefont
  {Aldegunde}, \citenamefont {Rivington}, \citenamefont {\.{Z}uchowski},\ and\
  \citenamefont {Hutson}}]{aldegunde:08}%
  \BibitemOpen
  \bibfield  {author} {\bibinfo {author} {\bibfnamefont {J.}~\bibnamefont
  {Aldegunde}}, \bibinfo {author} {\bibfnamefont {B.~A.}\ \bibnamefont
  {Rivington}}, \bibinfo {author} {\bibfnamefont {P.~S.}\ \bibnamefont
  {\.{Z}uchowski}}, \ and\ \bibinfo {author} {\bibfnamefont {J.~M.}\
  \bibnamefont {Hutson}},\ }\href {\doibase 10.1103/PhysRevA.78.033434}
  {\bibfield  {journal} {\bibinfo  {journal} {Phys. Rev. A}\ }\textbf {\bibinfo
  {volume} {78}},\ \bibinfo {pages} {033434} (\bibinfo {year}
  {2008})}\BibitemShut {NoStop}%
\bibitem [{Kot()}]{Kotochigova}%
  \BibitemOpen
  \href@noop {} {}\bibinfo {note} {S. Kotochigova, private communication,
  (2015).}\BibitemShut {Stop}%
\end{thebibliography}%


\setcounter{figure}{0}
\setcounter{equation}{0}
\setcounter{section}{0}

\clearpage
\onecolumngrid
\vspace{\columnsep}

\newcolumntype{Y}{>{\centering\arraybackslash}X}
\newcolumntype{Z}{>{\raggedleft\arraybackslash}X}

\newlength{\figwidth}
\setlength{\figwidth}{0.45\textwidth}

\newcommand{\NaK}{{$^{23}$Na$^{40}$K}}
\renewcommand{\thefigure}{S\arabic{figure}}
\renewcommand{\theHfigure}{Supplement.\thefigure}
\renewcommand{\theequation}{S\arabic{equation}}
\renewcommand{\thesection}{\arabic{section}}

\begin{center}
	\large{\textbf{Supplementary Information:\\ Resonant dipolar collisions of ultracold molecules induced by microwave dressing}}\\~\\
	
	\normalfont{Zoe Z. Yan$^1$, Jee Woo Park$^2$, Yiqi Ni$^1$, Huanqian Loh$^3$, Sebastian Will$^4$, Tijs Karman$^5$, and Martin Zwierlein$^1$\\
	\textit{$^1$MIT-Harvard Center for Ultracold Atoms, Research Laboratory of Electronics, and Department of Physics,
		Massachusetts Institute of Technology, Cambridge, Massachusetts 02139, USA\\
		$^2$Department of Physics, Pohang University of Science and Technology, Pohang 37673, Korea\\
		$^3$Department of Physics and Centre for Quantum Technologies, National University of Singapore, 117543, Singapore\\
		$^4$Department of Physics, Columbia University, New York 10027, USA\\
		$^5$ITAMP, Harvard-Smithsonian Center for Astrophysics, Cambridge, Massachusetts 02138, USA
	}%
}
\end{center}

\clearpage

\section{I. Microwave sweep for collision rate measurements\label{sec:expt}}
For the measurements of the collision rate, the microwave field is adiabatically swept from far off resonance to the final detuning. The field is generated by mixing a fixed microwave frequency source at $\omega_\mathrm{MW}\,{=}\,$\SI{5.573 379}{\giga\hertz} with a programmable radio-frequency source at $\omega_\mathrm{RF}$. The radio-frequency source is tuned so the higher-frequency sideband $\omega_\mathrm{MW}\,{+}\,\omega_\mathrm{RF}$ is initially $\delta_\mathrm{initial}$ away from the $\ket{g_1}\rightarrow\ket{f}$ resonance, ending at a detuning of $\delta_\mathrm{final}$ after a linear sweep lasting 4~ms. $\delta_\mathrm{initial}$ is typically \SI{12}{\kilo\hertz} below $\delta_\mathrm{final}$. The carrier frequency and the lower sideband are more than \SI{70}{\mega\hertz} detuned from any rotational transitions, and are not expected to play any role in the dynamics.

The radiation characteristics of the microwave antenna were empirically determined via microwave spectroscopy. Rotational transitions were driven to multiple hyperfine states in $J\,{=}\,1$, and the microwave absorption features were scaled to match the line strengths determined by a theoretical model of the single-molecule Hamiltonian~\cite{Will2016}. The resulting radiation has an estimated 50\% $\pi$, 25\% $\sigma^+$, and 25\% $\sigma^-$ character.

\section{II. Coupled-channels calculations}

Section~\ref{sec:monomerH}II\,A gives a brief description of the single molecule Hamiltonian, $\hat{H}^{(X)}$; see the supplement of Ref.~\cite{Karman2018a} for more details.
The Hamiltonian for the pair of colliding molecules is given by
\begin{align}
	\hat{H} = -\frac{\hbar^2}{2\mu} \frac{d^2}{dR^2}+ \frac{\hbar^2 \hat{L}^2}{2\mu R^2} + \hat{H}^{(A)} + \hat{H}^{(B)} + \hat{V}_\mathrm{dip-dip}(R).
\end{align}
The first two terms correspond to the radial and centrifugal parts of the relative kinetic energy.
The last term represents the interaction between the two molecules,
for which we use the dipole-dipole interaction.
This describes dipole-dipole interactions induced by microwave dressing,
but also the dominant interaction in the absence of microwaves:
the van der Waals (rotational dispersion) interaction, which arises in second order from dipole-dipole coupling to the first rotationally excited state.
The resulting adiabatic potential curves are analyzed in Sec.~\ref{sec:potentials}III.

The coupled-channels equations are solved numerically using the renormalized Numerov algorithm of Ref.~\cite{janssen:13}.
This method yields two linearly independent sets of solutions,
and subsequently any desired boundary condition can be imposed.
The boundary conditions chosen are the usual $S$-matrix boundary conditions at long range,
and a fully absorbing boundary condition at short range.
The short-range boundary condition is imposed at $R_\mathrm{min}=50$~$a_0$,
but the calculated rates are independent of where precisely the short-range boundary condition is imposed,
as long as $R_\mathrm{min}$ is small compared to the remaining length scales in the problem.
In the absence of microwaves, the dominant interaction is rotational dispersion which acts on a length scale $R_\mathrm{vdW}^{(\mathrm{rot})} \,{=}\,(2\mu C_6/\hbar^2)^{1/4} \,{=}\, 491~a_0$,
whereas resonant dressing induces dipole-dipole interactions for which $R_\mathrm{dip} \,{=}\, d_0^2/6 (4\pi\epsilon_0\hbar^2) \, {=}\ 1.1 \times 10^4~a_0$.
The main physical idea is that microwave dressing affects the interactions experienced by the ground-state molecules.
Inducing long-range dipole-dipole interactions and suppressing the $p$-wave centrifugal barrier increase the flux of molecules that reach short range,
which leads to increased loss of molecules that is measured experimentally.

The coupled-channels calculations use uncoupled basis sets of the form
\begin{align}
	|J_A m_{J_A}\rangle|I_{A_1} m_{I_{A_1}}\rangle|I_{A_2} m_{I_{A_2}}\rangle|J_B m_{J_B}\rangle|I_{B_1} m_{I_{B_1}}\rangle|I_{B_2} m_{I_{B_2}}\rangle|L M_L\rangle|N\rangle,
\end{align}
which describe the rotational state and nuclear spin for both molecules,
the relative angular momentum of the colliding molecules $L$,
and the microwave photon number $N$.
These functions are adapted to permutation of identical molecules as is described in Ref.~\cite{Karman2018a}.
The basis sets are truncated as follows:
We include functions with $J=0$ and $1$, $L=1,3,5$, and $N=N_0, N_0-1,N_0-2$.
The nuclear spin states included have $\mIK=-4+\Delta \mIK$ and $\mINa=3/2+\Delta \mINa$ with $|\Delta m|\le 2$. 

In this work, we calculate two-body loss rates by explicitly thermally averaging cross sections obtained from coupled-channels calculations.
This is described in Sec.~\ref{sec:rates}II\,B,
and differs from the simpler approach used in Ref.~\cite{Karman2018a}.

\subsection{A. Single-molecule Hamiltonian \label{sec:monomerH}}

The molecules are modeled as rigid rotors with a dipole moment.
The monomer Hamiltonian of molecule $X$ is thus given by
\begin{align}
	\hat{H}^{(X)} = B_\mathrm{rot} \hat{J}^2 + \hat{H}_\mathrm{MW}^{(X)} + \hat{H}_\mathrm{hyperfine}^{(X)}.
	\label{eq:monH}
\end{align}
The first term describes the rigid rotor's rotational kinetic energy, with rotational constant $B_\mathrm{rot}$.

The second term in Eq.~\eqref{eq:monH} represents the interaction with a microwave electric field
\begin{align}
	\hat{H}_\mathrm{MW}^{(X)} = -\frac{E_\mathrm{MW}}{\sqrt{N_0}} \left[ \hat{d}_\sigma^{(X)} \hat{a}_\sigma + \hat{d}_\sigma^{(X)\dagger} \hat{a}_\sigma^\dagger\right] + \hbar\omega  \hat{a}_\sigma^\dagger \hat{a}_\sigma.
\end{align}
Here, $a_\sigma^\dagger$ and $a_\sigma$ are creation and annihilation operators for photons with polarization $\sigma$ and angular frequency $\omega$.
The microwave electric field strength is given by $E_\mathrm{MW}=\hbar\Omega/2d_{g_1,f}$, and $N_0$ is the reference number of photons. $\Omega \equiv \Omega_R$ is the Rabi frequency.
The dipole operator has spherical components $\sigma=0,\pm1$ which are related to the Cartesian components by $\hat{d}^{(X)}_0 = \hat{d}^{(X)}_z$ and $\hat{d}^{(X)}_{\pm 1} = \mp \left( \hat{d}^{(X)}_x \pm i \hat{d}^{(X)}_y\right)/\sqrt{2}$, corresponding to polarizations $\pi$ and $\sigma^\pm$.
The actual polarization in the experiment is estimated from the relative strength of various microwave transitions.
This polarization is at an angle to the magnetic field direction, which breaks the cylindrical symmetry in the system.
To simplify the calculations, we include only the $\sigma^+$ polarization component that should be dominant close to the $\sigma^+$ resonance used for dressing.
As discussed in Sec.~\ref{sec:expt}I and the main text,
the effect of the remaining components is not negligible away from resonance,
and will result in a larger effective Rabi frequency and correspondingly faster losses for large detuning.
This increase of the scattering rate at large detuning is also found in the analytical Condon approximation,
discussed in Sec.~\ref{sec:condon}IV.

The last term in Eq.~\eqref{eq:monH} represents the hyperfine Hamiltonian
\cite{aldegunde:08},
\begin{align}
	\hat{H}_\mathrm{hyperfine}^{(X)} = \hat{H}_{eQq}^{(X_1)} + \hat{H}_{eQq}^{(X_2)} + c_1 \hat{i}^{(X_1)} \cdot \hat{J} + c_2 \hat{i}^{(X_2)} \cdot \hat{J} + H^{(X)}_3 + c_4 \hat{i}^{(X_1)} \cdot \hat{i}^{(X_2)} + \hat{H}^{(X)}_\mathrm{Zeeman}.
\end{align}
These describe the interaction between the nuclear quadrupole moment and the internal electric field gradient for both molecules,
spin-rotation couplings,
and direct and indirect dipole-dipole couplings.
These interactions are described in more detail in the Supplementary Information to Ref.~\cite{Karman2018a}.

Table~\ref{tab:molconst} summarizes the values of the molecular constants used in this work.

\setlength{\tabcolsep}{30pt}
\begin{table}
	\caption{
		Molecular constants of \NaK~used in this work, from Refs.~\cite{Park2015,aldegunde:17}.
		\label{tab:molconst}}
	\begin{tabular}{lr}
		\hline\hline
		\hline
		$B_\mathrm{rot}$     & 2.8217~GHz  \\
		$d_0$                & 2.72~Debye  \\
		$\alpha_2$	     & 435~$a_0^3$ \cite{Kotochigova}\\ 
		\\
		$i^{(1)}$            & 3/2         \\
		$i^{(2)}$            & 4           \\
		$(eQq)^{(1)} $       & -187~kHz    \\
		$(eQq)^{(2)} $       & 899~kHz     \\
		$c_1$                & 117.4~Hz    \\
		$c_2$                & -97~Hz      \\
		$c_3$                & 48.4~Hz     \\
		$c_4$                & -409~Hz     \\
		& \\
		$g_{\rm r}$          & 0.0253      \\
		$g_1$                & 1.477       \\
		$g_2$                & -0.324      \\
		\hline\hline
	\end{tabular}
\end{table}

\subsection{B. Cross sections and rates \label{sec:rates}}

The coupled-channels calculations performed here yield $S$-matrices for the combined set of short-range and long-range product channels.
From the long-range part, we compute cross sections for inelastic scattering from the initial channel, $i$, to other hyperfine states or field-dressed levels, $f$, as
\begin{align}
	\sigma_{i\rightarrow f}^{(\mathrm{inel})} = \frac{2\pi}{k^2} \sum_{L,M_L,L',M_L'} \left| T^{(\mathrm{LR})}_{f,L',M_L';i,L,M_L} \right|^2,
\end{align}
where the $T$-matrix is given by $\bm{T}=\bm{1}-\bm{S}$.
We also define a short-range capture cross section
\begin{align}
	\sigma^{(\mathrm{SR})} = \frac{2\pi}{k^2} \sum_{r,L,M_L} \left| T^{(\mathrm{SR})}_{r;i,L,M_L} \right|^2.
\end{align}
Here, $(\mathrm{SR})$ denotes the short-range capture part and the sum over $r$
extends over all adiabatic channels that are classically allowed at the capture
radius, ${R_\mathrm{min}=50~a_0}$.

In Ref.~\cite{Karman2018a}, scattering calculations were performed at a single energy $E=1~\mu$K, and the energy dependence was assumed from the threshold behavior for identical bosons.

Here, we consider collisions of identical fermions that undergo $p$-wave collisions.
Furthermore, the microwave dressing induces interactions that significantly suppress the $p$-wave centrifugal barrier.
Therefore, as a function of microwave dressing, the energy dependence of the cross sections varies between that expected for $p$-wave and barrierless collisions.
To account for the energy dependence, we calculate cross sections for 50 logarithmically spaced energies between 1~nK and 10~$\mu$K.
Thermal rate coefficients are then calculated by averaging the cross sections over a Maxwell-Boltzmann distribution,
\begin{align}
	\beta = \sqrt{\frac{8 k_{\rm B}T}{\pi \mu}} \frac{1}{(k_{\rm B}T)^2} \int_0^\infty \sigma(E) \exp\left(-\frac{E}{k_\mathrm{B}T}\right) E dE,
\end{align}
where the temperature is $T=560$~nK.

\section{III. Interaction potentials \label{sec:potentials}}
\subsection{A. Dipole-dipole interactions \label{sec:dipdip}}

Here, we consider the dipole-dipole interactions for dressing on resonance ($\delta=0$) for a $\sigma^+$ transition.
Initially, we will ignore hyperfine interactions for simplicity.
This is appropriate if $\hbar\Omega$ is much larger than typical hyperfine splittings.
Molecules are prepared in the field-dressed ground state, ${|-\rangle = ( |J\,{=}\,0,m_J\,{=}\,0,N_0\,{=}\,0\rangle - |1,1,-1\rangle ) / \sqrt{2}}$; see Fig.~1(a) of the main text.
This state has acquired a rotating dipole moment in the lab frame, and the first-order interaction arises from the time-averaged dipole-dipole interaction
\begin{align}
	V^{(1)}(\bm{R}) = \langle -- | \hat{V}_\mathrm{dip-dip}(\bm{R}) |--\rangle = \frac{d_0^2}{6R^3} P_2(\cos\theta),
	\label{eq:1odd}
\end{align}
where $\theta$ is the angle between the intermolecular axis and the lab-frame $z$ axis and $P$ is the Legendre polynomial.
This is the leading interaction for large $R$, and is an accurate approximation to the full interaction if the dipole-dipole coupling is weak compared to $\hbar\Omega$.

In reality, the hyperfine splittings are larger than the Rabi frequencies used here.
This means that other hyperfine states are energetically well separated from the initial state. This includes the ``spectator states'',
i.e. the $m_J$ components of the excited states not addressed by the microwaves. For the first-order interaction, the effect of hyperfine interactions is to reduce the effective dipole moment in Eq.~\eqref{eq:1odd} due to the different nuclear spin state decomposition in the rotational ground and excited state.

At short $R$, where the dipole-dipole interaction becomes larger than $\hbar\Omega$ and $k_\mathrm{B}T$, the molecular dipole moments quantize along the intermolecular axis
rather than along the microwaves' polarization direction.
The eigenstates of the dipole-dipole interaction are the body-fixed basis functions,
$\hat{\mathcal{R}} \left[ |0,0\rangle|1,K\rangle + |1,K\rangle|0,0\rangle\right]/\sqrt{2} \sqrt{\frac{2J+1}{4\pi}} D^{(J)\ast}_{M,K}(\hat{\mathcal{R}})$.
Here, $\hat{\mathcal{R}}$ is the rotation operator that transforms between the space and body-fixed frames,
and $D^{(J)}_{M,K}$ is a Wigner D-matrix element.
The good quantum numbers are the total angular momentum $J$, its space-fixed projection quantum number $M$,
and the body-referred projection $K$.
The dipole-dipole interactions are $+1/3d_0^2 R^{-3}$ for $K=\pm 1$ and $-2/3d_0^2R^{-3}$ for $K=0$.
These are called resonant dipole-dipole interactions as they arise from transition dipole moments for the resonant rotational excitation and de-excitation of the two molecules.
States with both molecules in the ground state or both in the excited state also exist but experience no dipole-dipole interaction.

Hyperfine interactions result in splittings between the different $m_J$ components in the rotationally excited states.
When large compared to the dipole-dipole interaction, the hyperfine interactions prevent the molecules from quantizing along the intermolecular axis, which requires mixing of different $m_J$ substates.

\subsection{B. Adiabatic Potential Curves \label{sec:adiabats}}

\begin{figure*}
	\begin{center}
		\includegraphics[width=\figwidth,clip]{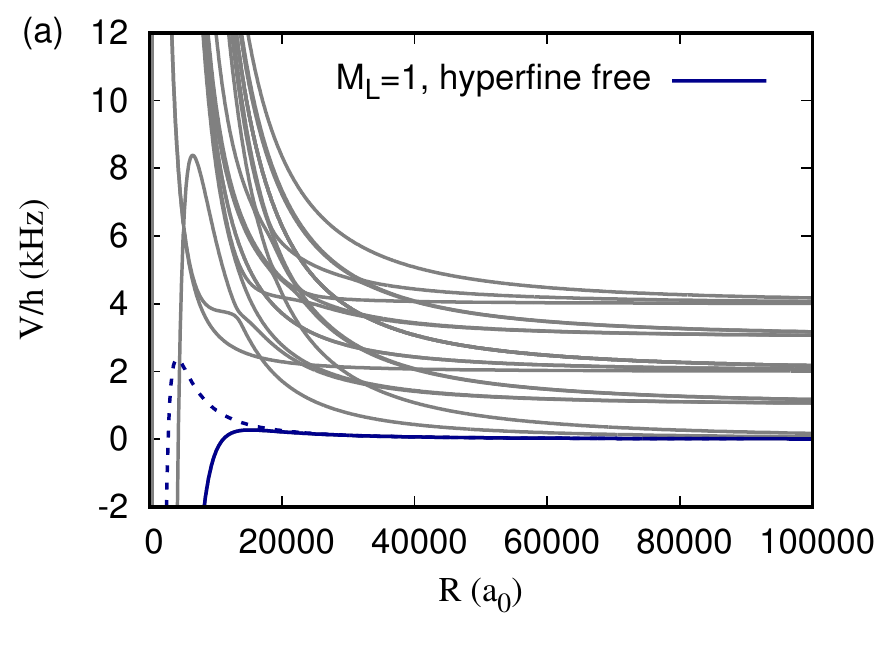}
		\includegraphics[width=\figwidth,clip]{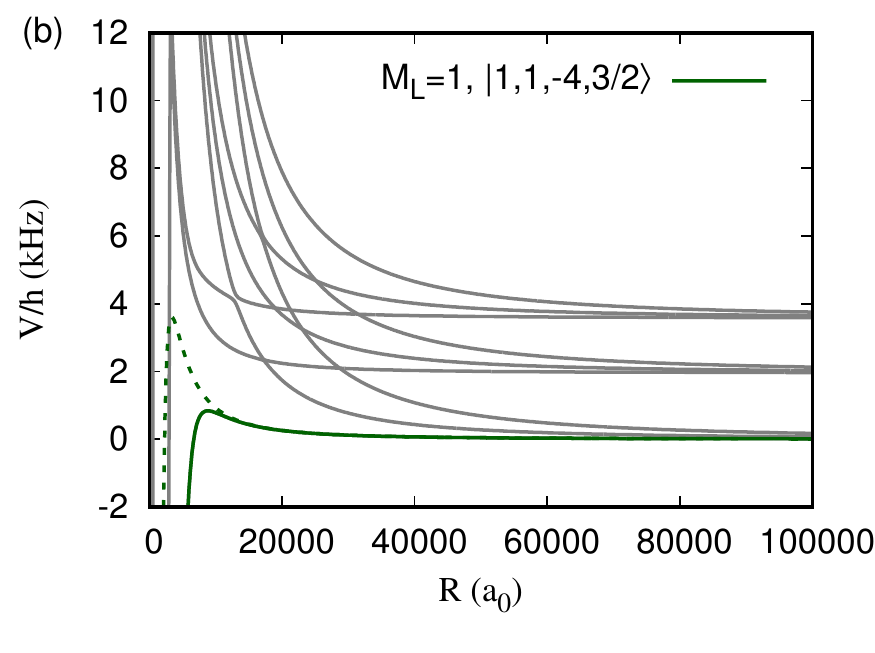}
		\includegraphics[width=\figwidth,clip]{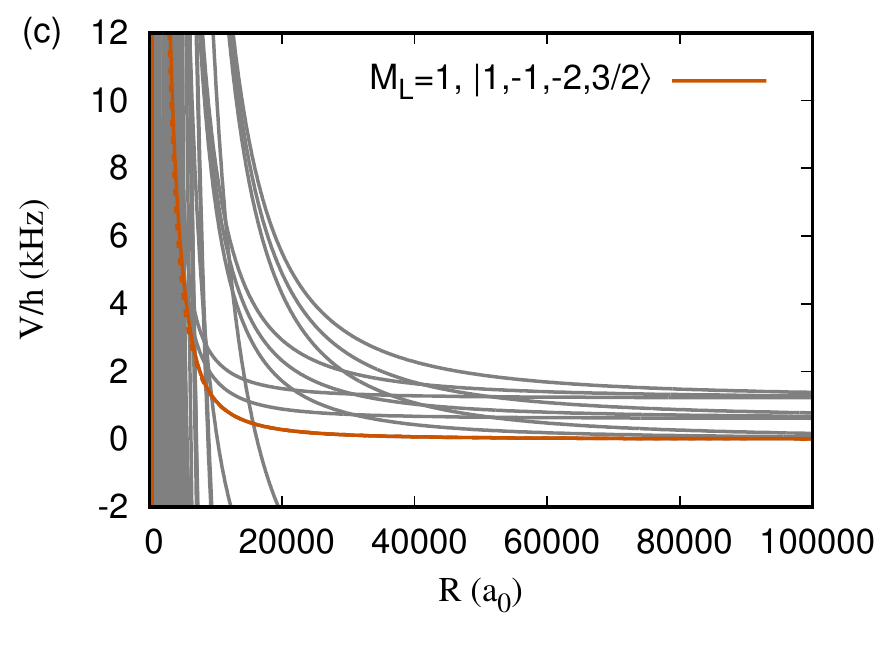}
		\includegraphics[width=\figwidth,clip]{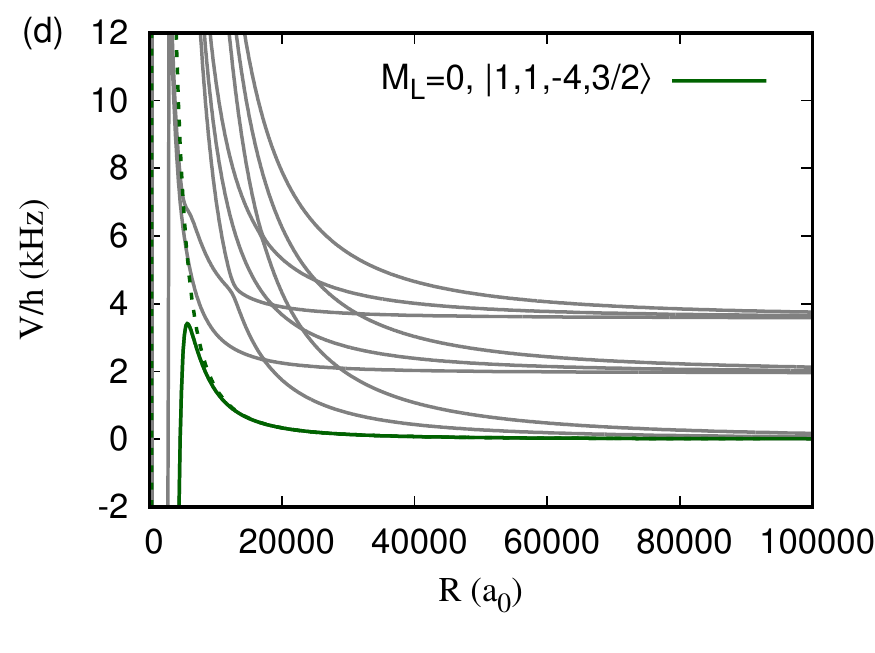}
		\caption{\label{fig:potentials}
			Adiabatic potential curves for resonant dressing. The lowest $p$-wave potential is highlighted in color,
			and the potential expected for first-order dipole-dipole interactions is shown as the dashed colored line,
			whereas the remaining adiabats are shown in gray.
			The different panels correspond to different hyperfine states and $M_L$ as indicated and discussed in the text, with the case of \textbf{(b)} being relevant to our experimental conditions.
		}
	\end{center}
\end{figure*}

\begin{figure*}
	\begin{center}
		\includegraphics[width=\figwidth,clip]{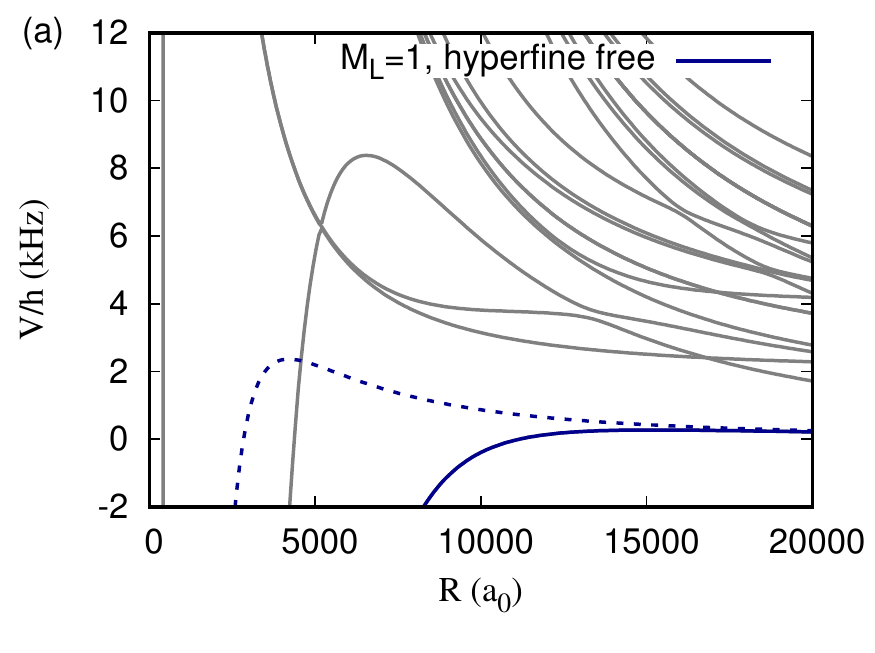}
		\includegraphics[width=\figwidth,clip]{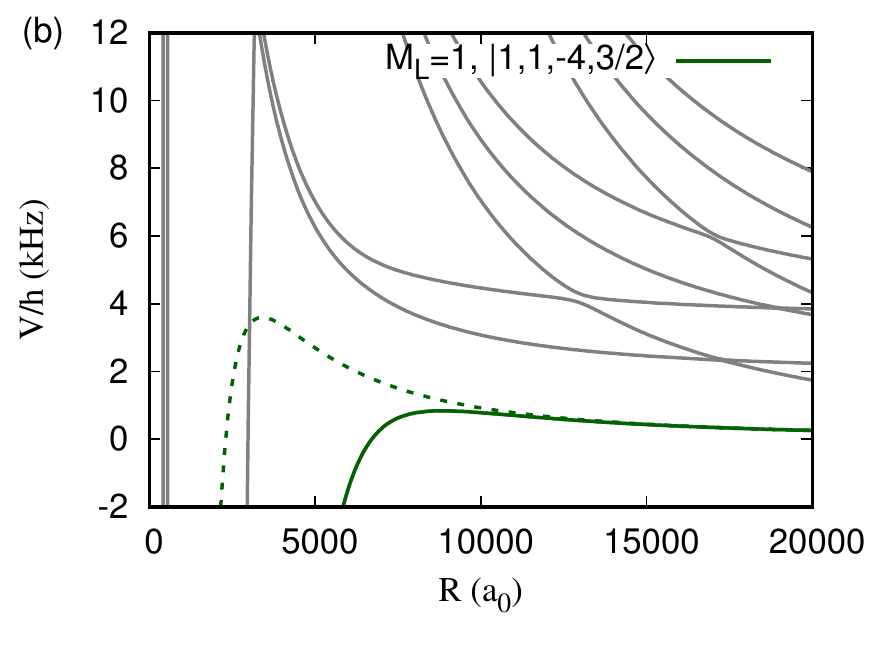}
		\includegraphics[width=\figwidth,clip]{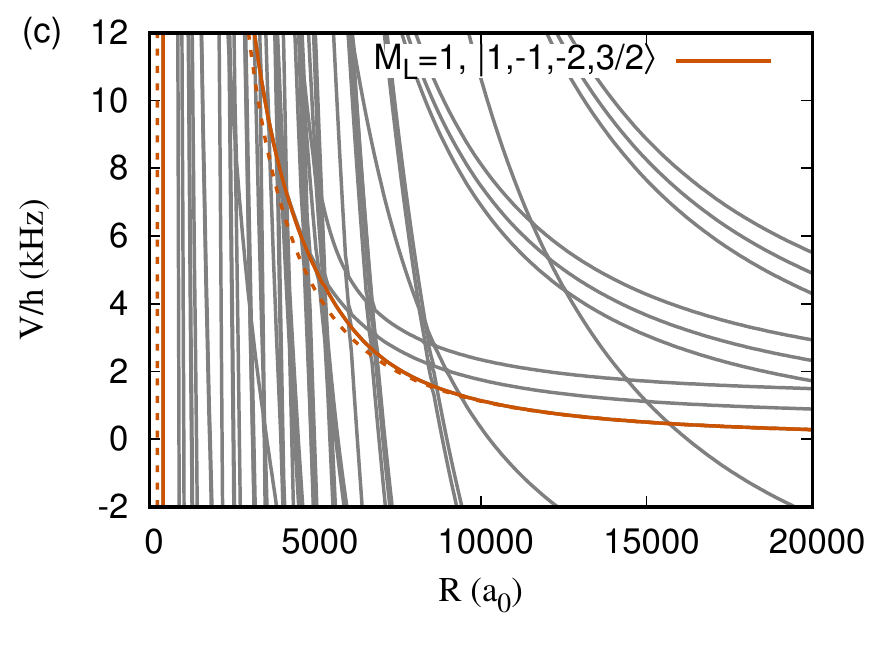}
		\includegraphics[width=\figwidth,clip]{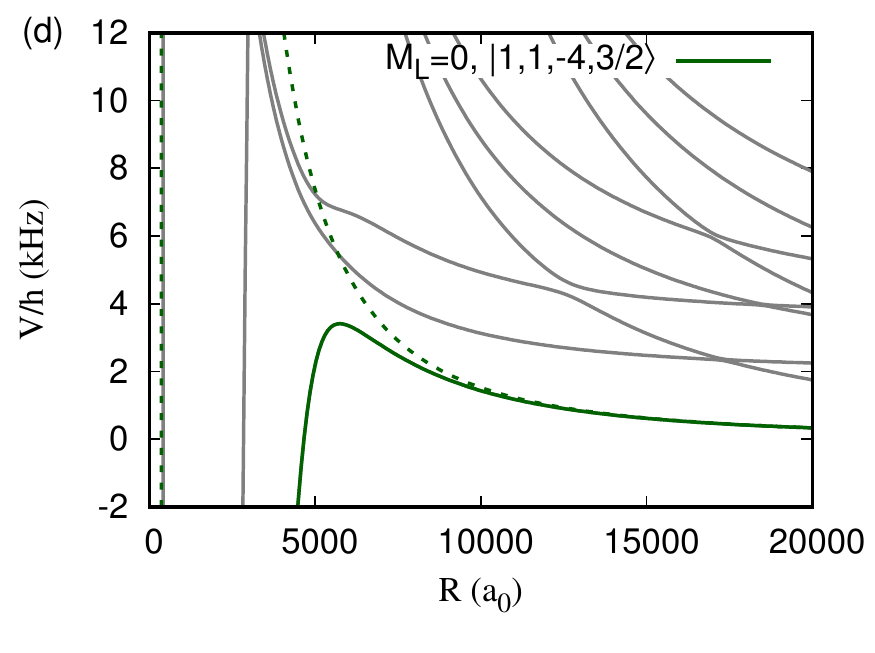}
		\caption{\label{fig:potentials_zoom}
			A version of Fig.~\ref{fig:potentials} showing detail at shorter distances.}
		
	\end{center}
\end{figure*}

Here, we analyze the adiabatic potential curves obtained by diagonalizing the total Hamiltonian excluding radial kinetic energy at fixed intermolecular distance, $R$.
The scattering calculations do not use the adiabatic representation, but it is used here as an interpretative tool.
Fig.~\ref{fig:potentials} shows plots of the potentials for resonant dressing ($\delta\,{=}\,0$).
Various adiabatic potential curves are shown as gray lines, whereas the lowest potential is highlighted in color.
The dashed colored line shows the potential expected for first-order dipole-dipole interactions,
i.e. the interaction between the space-fixed rotating dipoles induced by the microwave dressing.
The adiabats shown are obtained for resonant dressing with $\Omega/2\pi = 2$~kHz.
The experimental temperature of 560~nK corresponds to a kinetic energy of 12~kHz.
In the absence of microwaves, the interaction is dominated by rotational dispersion,
which leads to a $p$-wave centrifugal barrier of 200~kHz.
The four panels are discussed below.

Fig.~\ref{fig:potentials}(a) shows the adiabatic potential curves excluding hyperfine interactions.
This leads to five thresholds $|++\rangle$, $|+0\rangle$, $|00\rangle$ and $|+-\rangle$, $|-0\rangle$, and $|--\rangle$, which are split by half the Rabi frequency each.
Here, $\ket{0}$ represents a ``spectator state."
Each threshold shows multiple adiabatic potential curves, which correspond to the partial waves $L=1,3$, and $5$, included here.
Dipole-dipole couplings to the nearby field-dressed levels exist, and are given in Tab.~\ref{tab:tab}.
At long range, where the dipole-dipole interaction is weak compared to $\hbar\Omega$,
these couplings can be neglected and the interaction for the lowest adiabat is given by the first-order dipole-dipole interaction, $\langle{--}|\hat{V}_\mathrm{dip-dip}|{--}\rangle$.
At short range, where the dipole-dipole interaction is much stronger than $\hbar\Omega$,
the lowest adiabat approaches the attractive $K=0$ resonant dipole-dipole potential, $-2/3\;d_0^2/R^3$.
As discussed in the main text, for large detunings, the transition between these regimes occurs at the Condon point where this resonant dipole-dipole interaction is resonant with the detuning.
For large detuning, this means that the dipole-dipole interaction is also strong compared to the Rabi frequency, which then acts perturbatively, leading to a narrow avoided crossing.
On resonance, for which the potentials are shown in Fig.~\ref{fig:potentials}, the same transition occurs more gradually as the dipole-dipole interaction grows to be comparable to, and at some point dominant over, $\hbar\Omega$ as $R$ decreases when approaching short range.
This results in a significantly suppressed $p$-wave centrifugal barrier,
when compared to the barrier expected for first-order dipole-dipole interactions,
which is shown as the dashed line.

Fig.~\ref{fig:potentials}(b) shows the adiabatic potential curves including hyperfine interactions for dressing on the lowest hyperfine transition, $|0, 0, -4, 3/2\rangle \rightarrow |1,1,-4,3/2\rangle$.
A qualitative difference with the hyperfine-free case is that the thresholds corresponding to molecules in ``spectator states'', $|0\rangle$, appear to be missing.
These spectator states correspond to different hyperfine states, which are split by tens of kHz.
Hence, the lowest adiabat approaches the attractive $K=0$ resonant dipole-dipole potential only at shorter $R$.

Panel~\ref{fig:potentials}(c) shows the adiabatic potential curves for dressing with an excited hyperfine transition, $|1,-1,-2,3/2\rangle$.
The interactions within the nearly degenerate subset shown are weaker,
due to the smaller transition dipole moment.
Because here we are dressing with an excited hyperfine state, some hyperfine states occur below the initial threshold.
Many curve crossings occur near $R=5\,000~a_0$, where the Condon point otherwise occurs,
and the adiabatic potential does not approach an isolated $-2/3\;d_0^2/R^3$ curve in this case.
The curve crossings also lead to additional losses through inelastic transitions.
These effects can be seen in Fig.~4 of the main text, which shows calculated loss rates due to molecules reaching short range or inelastic transitions for various hyperfine states.
It should be noted that at short range, the density of adiabatic potential curves becomes very high and the interpretative power of the lowest adiabatic potential is lost.
The particular curve highlighted in solid orange is obtained by diabatically following the lowest initial adiabat inwards through narrowly avoided crossings.

Panels~\ref{fig:potentials}(a-c) show adiabatic potential curves for $M_L=1$, for which the dipole-dipole interaction induced by dressing on $\sigma^\pm$ transitions is attractive.
Panel~\ref{fig:potentials}(d) shows dressing for the same lowest $\sigma^+$ transition as is shown in \ref{fig:potentials}(b), but now for $M_L=0$.
In this case, the first-order dipole-dipole interaction is completely repulsive.
However, the lowest adiabatic curve contains only a small barrier that is well below the thermal kinetic energy of 12~kHz.
Hence, contrary to expectations from only first-order dipolar interactions, $M_L=0$ collisions can occur essentially without a barrier and add to the total scattering cross section by about one unitarity limited channel,
giving rise to the fast collisional loss observed in this work.

Fig.~\ref{fig:potentials_zoom} shows a zoomed-in version of Fig.~\ref{fig:potentials}.

\begin{figure*}
	\begin{center}
		\includegraphics[width=\figwidth,clip]{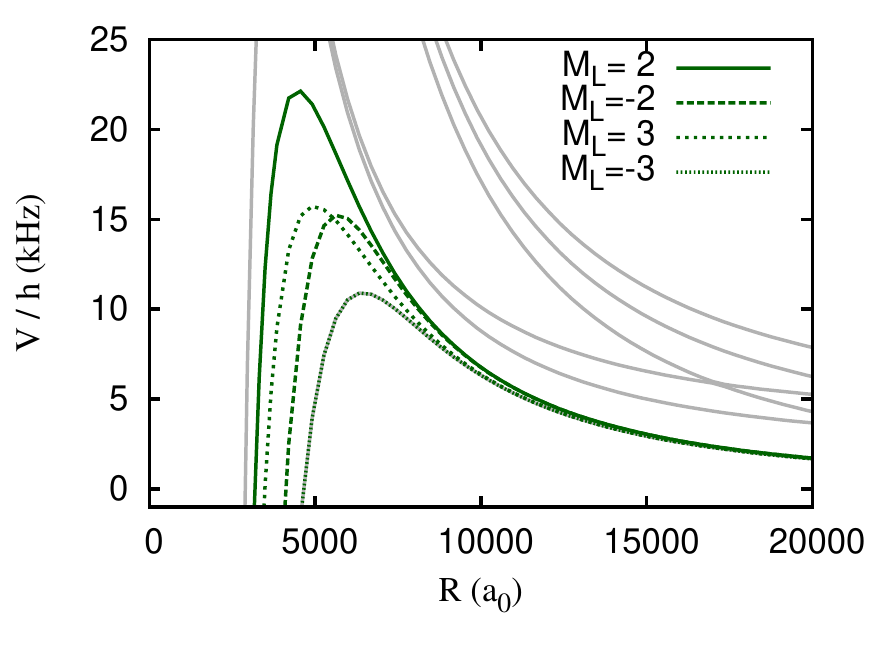}
		\caption{\label{fig:potL3}
			Centrifugal barriers for $f$-wave collisions, $L=3$ $|M_L|>1$, are highlighted in color for dressing on the lowest hyperfine transition, $|0, 0, 3/2, -4\rangle \rightarrow |1,1,-4,3/2\rangle$.
			Remaining adiabats for $M_L=-3$ are shown in gray for reference.
			The resulting barriers are comparable to $k_\mathrm{B}T/h = 12$~kHz, such that $f$-wave collisions contribute significantly to the loss.
		}
	\end{center}
\end{figure*}

Similarly, we find that the centrifugal barriers are significantly suppressed for $L=3$ and $|M_L|>1$.
The resulting $f$-wave centrifugal barriers, shown in Fig.~\ref{fig:potL3}, are comparable to $k_\mathrm{B}T$ and are suppressed to an order of magnitude below the $p$-wave centrifugal barrier in the absence of microwaves.

\begin{table}
	\caption{
		List of the states of the dimer where both molecules are in the field-dressed ground state, $|-\rangle$, excited state, $|+\rangle$, or one of the spectator states $|0\rangle \equiv |J=1,m_J = 0\rangle$ and $|\bar{0}\rangle \equiv |J=1,m_J = -1\rangle$.
		This lists dipole-dipole coupling to the ground state, $|--\rangle$, and energy with respect to this state.
		This table is valid on resonance, $\delta=0$, and does not include Zeeman or hyperfine interactions that increase the energy of the spectator states $|0\rangle$ and $|\bar{0}\rangle$.
		$C_{\ell,m}(\hat{R})$ is a Racah-normalized spherical harmonic depending on the lab-fixed polar angles of the intermolecular axis.
		\label{tab:tab}}
	\begin{tabularx}{\columnwidth}{lYr}
		\hline
		\hline
		state & $\langle -- | \hat{V}_\mathrm{dip-dip} | \mathrm{state} \rangle$ & $\Delta E$\\
		\hline
		$|++\rangle$                                                  & $-\frac{d_0^2}{6R^3} C_{2,0}(\hat{R})$            & $2\hbar\Omega$\\
		$\left(|\bar{0}-\rangle + |-\bar{0} \rangle\right)/\sqrt{2}$  & $\frac{d_0^2}{\sqrt{6}R^3} C_{2,2}(\hat{R})$      & $\frac{3}{2}\hbar\Omega$\\
		$\left(|0-\rangle + |-0 \rangle\right)/\sqrt{2}$              & $\frac{d_0^2}{\sqrt{12}R^3} C_{2,1}(\hat{R})$     & $\frac{3}{2}\hbar\Omega$\\
		$|+-\rangle$                                                  & 0                                                 & $\hbar\Omega$\\
		$|00\rangle$                                                  & 0                                                 & $\hbar\Omega$\\
		$\left(|\bar{0}-\rangle + |-\bar{0} \rangle\right)/\sqrt{2}$  & $-\frac{d_0^2}{\sqrt{6}R^3} C_{2,2}(\hat{R})$     & $\frac{1}{2}\hbar\Omega$\\
		$\left(|0-\rangle + |-0\rangle\right )/\sqrt{2}$              & $-\frac{d_0^2}{\sqrt{12}R^3} C_{2,1}(\hat{R})$    & $\frac{1}{2}\hbar\Omega$\\
		$|--\rangle$                                                  & $\frac{d_0^2}{6R^3} C_{2,0}(\hat{R})$             & 0 \\
		\hline
		\hline
	\end{tabularx}
\end{table}

\subsection{C. Simplified potentials}

\begin{figure*}
	\begin{center}
		\includegraphics[width=\textwidth,clip]{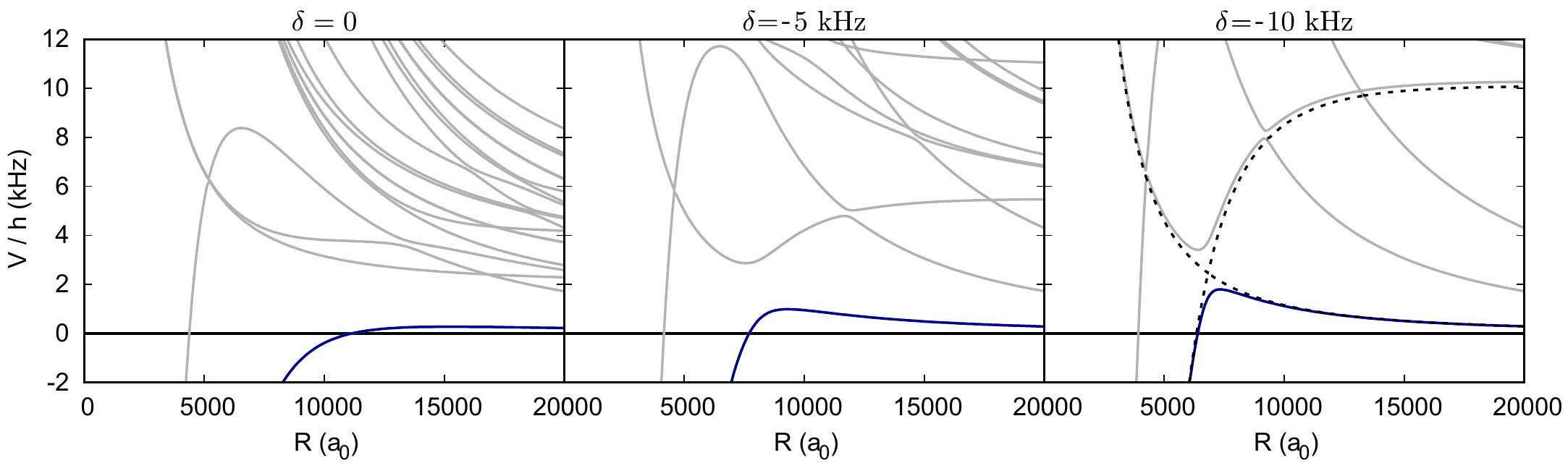}
		\caption{\label{fig:deltaprog}
			Evolution of potential curves for increasing red detuning and $M_L=1$, neglecting hyperfine interactions. The simplified two-state picture discussed in the main text emerges for detunings $|\delta| > \Omega_R$. The dashed lines for $\delta=-10$~kHz show the diabatic curves of Fig. 3(b) of the main text, given by the centrifugal potential for the incoming state and $-\hbar \delta - 2 d_0^2/3R^3$ for the attractive potential curve.
		}
	\end{center}
\end{figure*}

As we show in Fig.~\ref{fig:deltaprog}, the potential curves evolve with increasing red detuning towards the simpler two-state picture presented also in Fig.~3 of the main text. In this limit, a simple analytical treatment of the losses is possible.

To prepare for this, let us here derive simplified potentials obtained for fixed orientation of the intermolecular axis, $\hat{R}$, at an angle $\theta$ with respect to the laboratory $z$ axis,
in which frame the microwave polarization is defined.
We consider states for a pair of colliding molecules and the microwave field.
The bare ground state is denoted as $|j_A m_A, j_B m_B\rangle|N-N_0\rangle = |00,00\rangle|0\rangle$.
In addition we consider the following nearly degenerate states,
$|00,1\, -1\rangle_s|-1\rangle$, $|00,1\, 0\rangle_s|-1\rangle$, and $|00,1\, 1\rangle_s|-1\rangle$,
at energy $-\hbar\delta$,
and
$|1\, -1,1\, -1\rangle_s|-2\rangle$,
$|1\, -1,1\,  0\rangle_s|-2\rangle$, 
$|1\, -1,1\,  1\rangle_s|-2\rangle$, 
$|1\,  0,1\,  0\rangle_s|-2\rangle$,
$|1\,  0,1\,  1\rangle_s|-2\rangle$,
and $|1\, 1,1\,  1\rangle_s |-2\rangle$,
at energy $-2\hbar\delta$.
The subscript $s$ denotes symmetrization with respect to exchange of the two molecules.
A $\sigma^+$-polarized microwave field couples $|00\rangle|N\rangle$ and $|11\rangle|N-1\rangle$ at a strength parameterized by the Rabi frequency, $\Omega$.
Dipole-dipole interactions occur between states with the same microwave photon number and opposite parity for both molecular states.
Hence, within the nearly degenerate manifold, dipole-dipole coupling occurs only between the singly excited states.
Together, this leads to the following matrix representation of the Hamiltonian
\begin{align}
	\bm{H} =
	\begin{pmatrix}
		0        & 0        & 0        & \hbar\Omega/\sqrt{2}        & 0        & 0        & 0        & 0        & 0        & 0        \\ 
		0        & \frac{1}{3} \frac{d_0^2}{R^{3}} C_{2,0}(\hat{R}) -\hbar\delta        & \frac{1}{\sqrt{3}} \frac{d_0^2}{R^{3}} C_{2,1}(\hat{R})        & \sqrt{\frac{2}{3}} \frac{d_0^2}{R^{3}} C_{2,2}(\hat{R})        & 0        & 0        & \hbar\Omega/2        & 0        & 0        & 0        \\
		0        & -\frac{1}{\sqrt{3}} \frac{d_0^2}{R^{3}} C_{2,-1}(\hat{R})        & -\frac{2}{3} \frac{d_0^2}{R^{3}} C_{2,0}(\hat{R})-\hbar\delta        & -\frac{1}{\sqrt{3}} \frac{d_0^2}{R^{3}} C_{2,1}(\hat{R})        & 0        & 0        & 0        & 0        & \hbar\Omega/2        & 0        \\
		\hbar\Omega/\sqrt{2}        & \sqrt{\frac{2}{3}} \frac{d_0^2}{R^{3}} C_{2,-2}(\hat{R})        & \frac{1}{\sqrt{3}} \frac{d_0^2}{R^{3}} C_{2,-1}(\hat{R})        & \frac{1}{3} \frac{d_0^2}{R^{3}} C_{2,0}(\hat{R})-\hbar\delta        & 0        & 0        & 0        & 0        & 0        & \hbar\Omega/\sqrt{2}        \\
		0        & 0        & 0        & 0        & -2\hbar\delta        & 0        & 0        & 0        & 0        & 0        \\
		0        & 0        & 0        & 0        & 0        & -2\hbar\delta        & 0        & 0        & 0        & 0        \\
		0        & \hbar\Omega/2        & 0        & 0        & 0        & 0        & -2\hbar\delta        & 0        & 0        & 0        \\
		0        & 0        & 0        & 0        & 0        & 0        & 0        & -2\hbar\delta        & 0        & 0        \\
		0        & 0        & \hbar\Omega/2        & 0        & 0        & 0        & 0        & 0        & -2\hbar\delta        & 0        \\
		0        & 0        & 0        & \hbar\Omega/\sqrt{2}        & 0        & 0        & 0        & 0        & 0        & -2\hbar\delta        
	\end{pmatrix}.
\end{align}
Because there are no dipole-dipole interactions among the doubly excited states,
and the microwave field couples the ground state only to $m=+1$ excited states, doubly excited states with $m_A$ and $m_B \neq +1$ are completely uncoupled and can be removed from the basis.
This results in
\begin{align}
	\bm{H} =
	\begin{pmatrix}
		0        & 0        & 0        & \hbar\Omega/\sqrt{2}       & 0        & 0        & 0        \\
		0        & \frac{1}{3} \frac{d_0^2}{R^{3}} C_{2,0}(\hat{R}) -\hbar\delta        & \frac{1}{\sqrt{3}} \frac{d_0^2}{R^{3}} C_{2,1}(\hat{R})        & \sqrt{\frac{2}{3}} \frac{d_0^2}{R^{3}} C_{2,2}(\hat{R})        & \hbar\Omega/2      & 0        & 0        \\
		0        & -\frac{1}{\sqrt{3}} \frac{d_0^2}{R^{3}} C_{2,-1}(\hat{R})        & -\frac{2}{3} \frac{d_0^2}{R^{3}} C_{2,0}(\hat{R})-\hbar\delta        & -\frac{1}{\sqrt{3}} \frac{d_0^2}{R^{3}} C_{2,1}(\hat{R})        & 0      & \hbar\Omega/2        & 0        \\
		\hbar\Omega/\sqrt{2}        & \sqrt{\frac{2}{3}} \frac{d_0^2}{R^{3}} C_{2,-2}(\hat{R})        & \frac{1}{\sqrt{3}} \frac{d_0^2}{R^{3}} C_{2,-1}(\hat{R})        & \frac{1}{3} \frac{d_0^2}{R^{3}} C_{2,0}(\hat{R})-\hbar\delta        & 0        & 0        & \hbar\Omega/\sqrt{2}        \\
		0        & \hbar\Omega/2        & 0        & 0        & -2\hbar\delta        & 0        & 0        \\
		0        & 0        & \hbar\Omega/2       & 0         & 0        & -2\hbar\delta        & 0        \\
		0        & 0        & 0        & \hbar\Omega/\sqrt{2}         & 0        & 0        & -2\hbar\delta
	\end{pmatrix}
\end{align}
in the basis
$\{ |00,00\rangle, |00,1\,-1\rangle_s, |00,1\,0\rangle_s, |00,1\,+1\rangle_s, |1\, -1,1\,  1\rangle_s, |1\,  0,1\,  1\rangle_s, |1\, 1,1\,  1\rangle_s \}$
where the microwave photon number is implicit. 

For large detuning, $|\delta|\gg\Omega$, the Rabi coupling can be treated perturbatively,
and its effect will be significant where the resonant dipole-dipole interaction compensates for the detuning.
Therefore, the doubly excited channels can be ignored, leading to a $4\times 4$ problem.
The resonant dipole-dipole interaction is most conveniently described in the body-fixed frame,
which has the $z$ axis along the intermolecular frame.
Body-referred states with projection of angular momentum onto the intermolecular axis $K$ are given by
\begin{align}
	\hat{\mathcal{R}} |00,1K\rangle_s = \hat{\mathcal{R}} \left[ |00,1K\rangle + |1K,00\rangle \right]/\sqrt{2}.
\end{align}
These are eigenstates of the dipole-dipole interaction
\begin{align}
	\hat{V} \hat{\mathcal{R}} |00,1K\rangle_s = \hat{\mathcal{R}} |00,1K\rangle_s \times \begin{cases} +\frac{d^2}{3} R^{-3} & \text{for $K=\pm1$,} \\ -\frac{2d^2}{3} R^{-3} & \text{for $K=0$.} \end{cases}
\end{align}
In the body-fixed frame,
the effective Rabi coupling depends on the orientation of the intermolecular axis with respect to the lab frame,
in which the microwave polarization $\sigma$ is defined,
\begin{align}
	\hat{\mathcal{R}}^\dagger \hat{d}_\sigma \hat{\mathcal{R}} = \sum_{\kappa} D^{(1)\ast}_{\sigma,\kappa}(\hat{R}) \hat{d}_{\kappa}.
\end{align}
This leads to the Hamiltonian
\begin{align}
	\bm{H} =
	\begin{pmatrix}
		0                           & \hbar\Omega\sqrt{2} D^{(1)\ast}_{1,+1}(\hat{R}) & \hbar\Omega\sqrt{2} D^{(1)\ast}_{1,0}(\hat{R})  &  \hbar\Omega\sqrt{2} D^{(1)\ast}_{1,-1}(\hat{R}) \\
		\hbar\Omega\sqrt{2} D^{(1)}_{1, 1}(\hat{R}) & \frac{d_0^2}{3}R^{-3}-\hbar\delta  & 0                                     &  0 \\
		\hbar\Omega\sqrt{2} D^{(1)}_{1, 0}(\hat{R}) & 0                                  & -\frac{2d_0^2}{3}R^{-3}-\hbar\delta   &  0 \\
		\hbar\Omega\sqrt{2} D^{(1)}_{1,-1}(\hat{R}) & 0                                  & 0                                     &  \frac{d_0^2}{3}R^{-3}-\hbar\delta 
	\end{pmatrix}
\end{align}
The $K=\pm 1$ states experience repulsive resonant dipole-dipole interactions,
and at each fixed $\theta$ one ``bright" (microwave coupled) and one ``dark" (microwave uncoupled) state combination can be found.
The $K{=}0$ excited state experiences attractive resonant dipole-dipole interactions,
and the Rabi coupling to this state is most significant at the Condon point, $R_\mathrm{C} = (-2d_0^2/3\hbar\delta)^{1/3}$, where it crosses the flat ground-state potential, leading to an avoided crossing.


\section{IV. Condon approximation \label{sec:condon}}

Here, we consider analytically the loss rate for large detuning $\delta\gg\Omega$ and excluding hyperfine interactions.
The relevant potential curves are discussed in the main text and Fig.~3(b).
The ground state, $|g\rangle = |00,00\rangle|L M_L\rangle$, experiences essentially no dipole-dipole interactions,
and its potential is given by the centrifugal term and the background vdW interaction.
This potential is crossed by the excited state,
\begin{align}
	|f, J M \rangle = \hat{\mathcal{R}} \left[ |0,0\rangle|1,0\rangle + |1,0\rangle|0,0\rangle\right]/\sqrt{2} \sqrt{\frac{2J+1}{4\pi}} D^{(J)\ast}_{M,0}(\hat{\mathcal{R}}).
\end{align}
The excited state experiences attractive resonant dipole-dipole interactions and the potential is given by {$-\hbar\delta - 2/3\; d_0^2/R^3$}.
These potentials cross at the Condon point
\begin{align}
	R_\mathrm{C} = \left(\frac{2 d_0^2}{3 \hbar\delta} \right)^{1/3},
\end{align}
neglecting the ground state potential.
Finally, the ground and excited states are coupled by the microwave field
\begin{align}
	\langle f, J M | \hat{H}_\mathrm{MW} |g\rangle|L M_L\rangle &= \frac{\hbar\Omega \sqrt{3}}{2 d_0} \langle f, J M | \hat{d}_{\sigma} |g\rangle|L M_L\rangle \nonumber \\
	&= \hbar\Omega/\sqrt{2} \sqrt{\frac{2L+1}{2J+1}} \langle L M_L 1 \sigma | J M \rangle\langle L 0 1 0 | J 0 \rangle,
	\label{eq:Omangular}
\end{align}
where $\sigma = +1$, 0, or $-1$ for $\sigma^+$, $\pi$, or $\sigma^-$ polarized microwaves.

We assume losses occur with unit probability on the attractive excited state potential, where the molecules can reach short range,
and that non-adiabatic transitions to the excited state occur only at the Condon point \cite{Julienne1996,burnett1996laser,boisseau2000reflection}.
The loss rate is given by
\begin{align}
	\beta = \frac{2\pi\hbar}{\mu} \sum_{L,M_L,J,M} \left\langle \frac{1}{k} |S_{f, J M ;g,LM_L}|^2 \right\rangle,
\end{align}
where $k$ is the incoming wavenumber,
the angular brackets indicate an ensemble average,
and the squared $S$-matrix element is given by
\begin{align}
	|S_{i;j}|^2 = 4\pi^2 |\langle \Psi_i | \hat{H}_\mathrm{MW} | \Psi_j \rangle|^2,
\end{align}
where $|\Psi_i\rangle$ denotes an \emph{energy}-normalized scattering state in channel $i$.
The angular part of the matrix element above is given in Eq.~\eqref{eq:Omangular}.
The square of Eq.~\eqref{eq:Omangular} summed over $J$, $M$, and $M_L$ amounts to $\Omega^2/2$, regardless of the polarization of the microwave field, $\sigma$.
Hence, we obtain for the loss rate coefficient
\begin{align}
	\beta = \frac{4\pi^3 \Omega^2 \hbar}{\mu} \sum_{L} \left\langle \frac{1}{k} |\langle \psi_e | \psi_g \rangle|^2 \right\rangle.
\end{align}
The Franck-Condon factor can be computed in the reflection approximation \cite{Julienne1996,burnett1996laser,boisseau2000reflection},
\begin{align}
	|\langle \psi_e | \psi_g \rangle|^2 = |\psi_{g}(R_\mathrm{C})|^2 / D_\mathrm{C},
\end{align}
and depends only on the ground state wavefunction and the difference in potential slope at the Condon point
\begin{align}
	D_\mathrm{C} = \left| \frac{\mathrm{d} \left(V_g-V_e\right)}{\rm dR} \right|_{R_\mathrm{C}} = 2d_0^2 R_\mathrm{C}^{-4},
\end{align}
where we have neglected the background vdW interaction and the difference in centrifugal kinetic energy.
Neglecting the background vdW interaction, the energy-normalized ground state radial wavefunction is given by
\begin{align}
	\psi_{g}(R) = R \, j_l(k R) \sqrt{\frac{2 \mu k}{\pi}}
\end{align}
where $j_l$ is the spherical Bessel function of the first kind.
Combining the above, we obtain the loss rate given in Eq.~(1) of the main text,
and repeated here for completeness
\begin{align}
	\beta &= 4 \pi^2 \hbar \Omega_{\rm{R}}^2 R_\mathrm{C}^6 \langle j_l(k R_\mathrm{C})^2 \rangle / d_0^2 \nonumber \\
	&= \frac{16\pi^2}{9\hbar} \langle j_l(k R_\mathrm{C})^2 \rangle \, d_0^2 \, \left(\frac{\Omega_{\rm{R}}}{\delta}\right)^2
	\label{eq:Condon}
\end{align}
This expression does not include short-range loss of flux remaining in the ground-state potential,
\emph{i.e.}, the universal $p$-wave loss,
which has been added in Fig.~2(c) of the main text.

The loss rate in the Condon approximation, Eq.~\eqref{eq:Condon}, is independent of the polarization of the microwave field.
Therefore, we can account for the 50~\% $\pi$ and 25~\% $\sigma^-$ polarization components simply by increasing the effective Rabi frequency fourfold.
The resulting loss rate is shown as the red-dashed line in Fig.~2(c) of the main text.

\end{document}